\documentclass[useAMS,twocolumn,usegraphicx,usenatbib]{mn2e}
\usepackage{amssymb}
\usepackage{textcomp}
\usepackage{graphicx}
\usepackage{color}
\usepackage[T1]{fontenc}
\usepackage{times}

\newcommand{\rvir}{r_{\rm vir}}
\newcommand{\zlow}{z_{\rm low}}
\newcommand{\zhigh}{z_{\rm high}}
\newcommand{\tdAmin}{t_{\delta A}^{\rm min}}
\newcommand{\tdAmax}{t_{\delta A}^{\rm max}}
\newcommand{\tfmin}{t_{f}^{\rm min}}
\newcommand{\tfmax}{t_{f}^{\rm max}}

\newcommand{\klarge}{K_{15}}
\newcommand{\kinterm}{K_{10}}
\newcommand{\kshort}{K_{7.5}}

\title[Lagrangian Volume Deformations]{Lagrangian Volume Deformations around Simulated Galaxies}
\author[S. Robles, R. Dom\'{\i}nguez-Tenreiro, J. O\~norbe and F.~J. Mart\'{\i}nez-Serrano]{S. Robles$^{1}$\thanks{E-mail:
sandra.robles@uam.es (SR)}, R. Dom{\'{\i}}nguez-Tenreiro$^{1,2}$, J. O\~norbe$^{3}$ and F.J. Mart\'{\i}nez-Serrano$^{4,5}$\\
$^{1}$Departamento de F\'{\i}sica Te\'{o}rica, Universidad Aut\'{o}noma de Madrid, E-28049 Cantoblanco, Madrid, Spain\\
$^{2}$Astro-UAM, UAM, Unidad Asociada CSIC\\
$^{3}$Max-Planck-Institut f\"ur Astronomie, K\"onigstuhl 17, D-69117, Heidelberg, Germany  
\\
$^{4}$Dept. de F\'{i}sica y A.C., Universidad Miguel Hern\'{a}ndez, E-03202 Elche, Spain\\
$^{5}$Next Limit Dynamics SL, E-28048, Madrid, Spain
}
\begin{document}

\date{Accepted 2015 April 12. Received 2015 March 30; in original form 2014 October 20}

\pagerange{\pageref{firstpage}--\pageref{lastpage}} \pubyear{2015}

\maketitle

\label{firstpage}

\begin{abstract}
 We present a detailed analysis of the local evolution of 206 Lagrangian Volumes (LVs)  
selected at high redshift 
around galaxy seeds, identified in a large-volume $\Lambda$ cold dark matter ($\Lambda$CDM) hydrodynamical simulation.
 The LVs have a mass range of $1 - 1500 \times 10^{10} M_\odot$.
We follow the dynamical evolution of the density field inside
these initially spherical LVs from $z=10$ up to $\zlow = 0.05$,
witnessing highly non-linear, anisotropic
mass rearrangements within them, leading to the emergence of the {\it local} cosmic web (CW).
These mass arrangements have been analysed
in terms of the reduced inertia tensor $I_{ij}^r$, focusing 
on the evolution of the principal axes of inertia and their corresponding eigendirections, and
paying particular attention to the times when the evolution of these two structural elements declines.
In addition, mass and component effects along this process have also been investigated.
We have found that deformations are led by dark matter dynamics and they transform most of the initially spherical LVs into prolate shapes, 
i.e. filamentary structures. 
An analysis of the individual freezing-out time distributions for shapes and eigendirections shows 
that first most of the LVs fix their three axes of symmetry (like a skeleton) early on, 
while accretion flows towards them still continue. Very remarkably, 
we have found that more massive LVs fix their skeleton earlier on than less massive ones. 
We briefly discuss the astrophysical implications our findings could have, including the galaxy mass-morphology relation
and the effects on the galaxy-galaxy merger parameter space, among others.

\end{abstract}


\section{Introduction}
\label{Intro}

Over the last few decades, galaxy surveys such as the Two-degree-Field Galaxy Redshift Survey 
\citep[2dFGRS;][]{Colless:2001}, the 
Sloan Digital Sky Survey 
\citep[SDSS; e.g.][]{Tegmark:2004}, the Two-Micron All-Sky Survey 
\citep[2MASS;][]{Huchra:2005} and 
the 6dFGS 
\citep{Jones:2004} have revealed that galaxies gather in an  
intricate network, the so-called cosmic web 
\citep*[CW, after][]{Bond:1996}, made of filaments, walls, nodes 
which surround vast empty regions, the voids \citep{Zeldovich:1970,Shandarin:1989}. 
These structures can be found on scales from a few to hundreds of megaparsecs and include huge flat structures like the 
Great Wall \citep{Geller:1989} and the SDSS Great Wall \citep{Gott:2005}, the largest known structure in the local  
Universe, with a size larger than $400 h^{-1}$ Mpc and enormous empty regions like the Bo\"{o}tes void \citep{Kirshner:1981,Kirshner:1987}.

These results have been complemented by mappings of the dark matter (DM) spatial distribution through weak lensing observations like  
the Hubble Space Telescope Cosmic Evolution Survey 
\citep[COSMOS;][]{Massey:2007} and recent results from the 
Canada--France--Hawaii Telescope Lensing Survey 
\citep[CFHTLenS;][]{VanWaerbeke:2013}.

Summing up, analyses of the current large scale distribution of  galaxies and mass  show that both  are 
 hierarchically organised into a highly interconnected network, 
displaying a wealth of structures and substructures over a huge range of densities and scales.
This web can be understood as the main feature of the anisotropical nature of gravitational collapse \citep{Peebles:1980},
as well as of its intrinsic hierarchical character, and in fact it is the main dynamical engine responsible
for structure formation in the Universe  \citep{Sheth:2004,ShethVdWeygaert:2004,Shen:2006}, including galaxy scales \citep{DT:2011}.   

According to the standard model of cosmology, large-scale structures observed in the Universe today are seeded by 
infinitesimal primordial density and velocity perturbations. The physical processes underlying their dynamical development
until the CW emergence can be explained by theories and models on the gravitational instability,
later on corroborated by a profusion of cosmological simulations, the first of them purely  $N$-body simulations 
\citep[see e.g.,][]{Yepes:1992,Jenkins:1998,Pogosyan:1998,Colberg:2005,Springel:2005,Dolag:2006},
while recent ones include baryons and stellar physics too \citep[see e.g.,][]{DT:2011,Metuki:2014}. 

Indeed, the advanced non-linear stages of gravitational instability are described by the Adhesion Model  
(AM; see \citealt{Gurbatov:1984}; \citealt{Gurbatov:1989}; \citealt{Shandarin:1989}; \citealt{Gurbatov:1991}, \citealt{Vergassola:1994} and 
\citealt{Gurbatov:2012}, for a recent review),
an extension of the  popular non-linear Zeldovich Approximation \citep[hereafter ZA; see][]{Zeldovich:1970}. 
 In comoving coordinates the ZA can be expressed as a mapping from the 
Lagrangian space (the space of initial conditions $\vec{q}$) into the Eulerian space
(real  space) described as a translation by a generalised irrotational velocity-like vector
(the displacement field $\vec{s}(\vec{q})$) times the linear density growth factor  $D_{+}(t)$, where
the displacement can be written as a scalar potential gradient $\vec{s}(\vec{q}) = - \vec{\nabla}_q \Psi (\vec{q})$.
This approximation allows us to predict where singularities (locations with
infinite density) will appear as cosmic evolution proceeds (i.e., the $\vec{q}$ points where
the map has a vanishing determinant of the Jacobian matrix) and how they evolve into a sequence of 
caustics in real space. 
In this way, the ZA correctly but roughly describes the emergence of multistream flow regions, 
caustics and the structural skeleton of the CW
\citep*{Doroshkevich:1973,Buchert:1989,Buchert:1992,Shandarin:1989,Coles:1993,Melotta:1994,Melottb:1994,Melott:1995,Sahni:1995,Yoshisato:1998,Yoshisato:2006}.

It is well known, however, that the ZA is
not applicable once a substantial fraction of the mass elements are contained in
multistream regions, 
because it predicts that caustics thicken and vanish due to multistreaming soon after their
formation. One way of overcoming this issue is to introduce a small diffusion term in Zeldovich
momentum equation, in such a way that it has an effect only when and where particle
crossings are about to take place. This can be accomplished by introducing a non-zero
viscosity, $\nu$, and then taking the limit $\nu \rightarrow  0$: this is 
the AM, whose main advantage is that the momentum equation
looks like the Burgers' equation \citep{Burgers:1974} in the same limit,
and hence its analytical solutions are known.
A physically motivated derivation of the AM  can be found in
   \citet{Buchert:1998,Buchert:1999,Buchert:2005}. 

The AM implies that, at a given scale, walls, filaments and nodes (i.e., the cosmic web
elements) are successively formed, and then they vanish due to mass piling-up around nodes, 
to where mass elements travel through walls and 
filaments\footnote{Recently confirmed in detail through CW element identification in large volume $N$-body simulations by \citet{Cautun:2014}.}.  
Meanwhile, the same web elements
emerge at larger and larger scales, and are erased at these scales after some time.
Therefore, the AM  conveniently describes both  the
 anisotropic nature of gravitational collapse and the hierarchical nature of the process. 
In addition, the AM indicates that
the advanced stages of non-linear evolution act as a kind of smoothing procedure on different scales, 
by wiping mass accumulations off walls and filaments, first at small
scales and later on at successively larger ones, 
to the advantage  of nodes. Another implication of the AM is that node
centres (protohaloes at high $z$)  lie  on the former  filaments at any $z$.

A very interesting achievement of the AM is that the first successful reduction of the
cosmic large scale structure to a geometrical skeleton was done in this approximation 
\citep{Gurbatov:1989,Kofman:1990,Gurbatov:2012}, see also \citet{Hidding:2014}.
Later on \citet{Novikov:2006,Sousbiea:2008,Sousbieb:2008,Sousbie:2009,Sousbie:2011,SousbiePichon:2011,AragonCalvoa:2010} and 
\citet{AragonCalvob:2010} 
also discussed the skeleton or spine of large-scale structures  from purely topological
constructions in a given density field.  

Recently, a growing interest  
to identify and analyse elements of the CW in $N$-body simulations, as well as 
in galaxy catalogues, has led to the development of different  mathematical tools 
\citep{Stoica:2005,AragonCalvoa:2007,AragonCalvob:2007,AragonCalvob:2010,Hahna:2007,Hahnb:2007,Platen:2007,Stoica:2007,ForeroRomero:2009,Wu:2009,AragonCalvoa:2010,Bonda:2010,Bondb:2010,Genovese:2010,Gonzalez:2010,Jones:2010,Stoica:2010,Hoffman:2012,Cautun:2013,Tempel:2014}. 
These methods and algorithms are motivated by 
the study of the influence of large scale structures on galaxy formation 
\citep{Altay:2006,AragonCalvob:2007,Hahna:2007,Hahnb:2007,Paz:2008,Hahn:2009,Zhang:2009,Godlowski:2011,Codis:2012,Libeskind:2012,Libeskind:2013,AragonCalvo:2014,Metuki:2014}.
In a recent paper, \citet{Cautun:2014} have investigated the evolution of the CW from cosmological simulations,
focusing on the global evolution of their morphological components and their halo content.

From a dynamical point of view, \citet{Hidding:2014} go a step further by establishing
the link between the skeleton or spine of the CW, as described by the previous methods, and the development
of the density field. In fact, they describe for the first time
the details of caustic emergence as cosmic evolution proceeds. Their main result
is to show that all dynamical processes related to caustics happen at locations
placed near a set
of critical lines in Lagrangian space, that, when projected onto the Eulerian space, 
imply an increasing degree of connectedness
among initially disjoint mass accumulations in  walls or filaments, until a percolated structure forms,
i.e., the spine or skeleton of the large scale mass distribution.  
These authors compare their results with two dimensional $N$-body simulations. Note that,
due to the complexity of the problem,
they first work in two dimensional spaces, where caustic emergence
and percolation are described. Nevertheless, 
they expect no important qualitative differences
when three-dimensional spaces are considered instead. 

As we can see, in the last years different methods to quantify the cosmic web structure, classify its elements and study
its emergence and evolution have been developed and applied. However,
a detailed analysis of the {\it local} development of the density field around 
galaxy hosting haloes is still missing.
This is of major importance because of its close connection to the problem of galaxy formation,
in which case the effects of including gas processes need to be considered too.
It is worth noting that neither the ZA nor the AM include  gas effects in their
description of CW dynamics.

This analysis should first answer to the simplest questions related  to {\it local} shape deformation and 
spine emergence and the orientation of its main directions or symmetry axes around galaxy-to-be objects.
Besides, the very nature of these {\it local} processes, there are other interesting, simple, not-yet-elucidated 
related issues. For instance the characterisation of the times when deformation stops and orientation 
gets frozen, 
whether or not this local web evolution is mass dependent (i.e., the mass
of the halo-to-be)  or not, and if different components (DM, hot gas, cold baryons)
evolve in a similar way or there is a component segregation.   
We do not have at our disposal an analytical tool to perform such analyses, in consequence we   
need to resort to numerical simulations.

In order to answer these questions,
in this paper we investigate the impact of the local features of the Hubble flow imprinted on the deformation
  of initially spherical Lagrangian volumes (LVs) and the spine emergence, from high to low redshift. As known from
previous studies, the local Hubble flow is neither homogeneous nor isotropic, 
on the contrary, it contains shear terms (and small-scale vorticity at its most advanced stages)
 that distort cosmological structures.
We use cosmological hydrodynamical simulations to study the deformations of a sample of LVs through their reduced inertia tensor at different redshifts, 
which allows us to describe in a quantitative way the LV shape deformation and evolution, 
along with that of their symmetry axes. 
We analyse every component separately, that is, we compute the  reduced inertia tensor 
for DM, cold and hot baryons.

This paper is organised as follows. In $\S$\ref{sec:methods}, we outline the simulation method and the algorithms used to 
study the deformations of LVs.  
A brief summary on the ZA, the CW emergence in 2D and the AM is given in
$\S$\ref{UnderEvol}, where some of their implications, useful in this paper,
 are also addressed. 
Some relevant details of the
highly non-linear stages of gravitational instability, beyond  the ZA or the AM
are summarised in $\S$\ref{FurtherEvol}, to help to understand how our results about the LV evolution can 
be explained in the light of these models.  In $\S$\ref{EigenEvol},  the LV evolution   
is investigated in terms of the reduced inertia tensor eigenvectors, delaying the analysis in terms of its eigenvalues 
to the next section, $\S$\ref{sec:results}, focused on the mass and component effects and 
on the shape evolution of the selected LVs.
In $\S$\ref{sec:Percola} we study the freezing-out of eigendirections and shapes, presenting 
the distribution of the corresponding freezing-out times and looking for mass effects. 
Possible scale effects on the previous results are discussed in 
$\S$\ref{subsec:scaleeffects}. 
 Finally, we present our summary, conclusions and discussion in $\S$\ref{sec:conclusions}.

\section[]{Simulations and Methods}
\label{sec:methods}

\subsection{Simulations}
\label{sec:simul}

The simulations analysed here have been run under the GALFOBS I and II projects.
The GALFOBS (Galaxy Formation at Different Epochs and in Different Environments: Comparison with Observational Data) project aims to study 
the generic statistical properties of galaxies in various environments and at different cosmological epochs.
This project was a DEISA Extreme Computing Initiative (DECI)\footnote{ 
The DEISA Extreme Computing Initiative was launched in May 2005 by the DEISA Consortium, as a way to enhance its impact on science and 
technology}. 
GALFOBS I was run at LRZ (Leibniz-Rechenzentrum) Munich, as a European project.
Its continuation, GALFOBS II, was run at the Barcelona Supercomputing Centre, Spain.

All the runs were performed using P-DEVA, 
the parallelised version of the DEVA code \citep{Serna:2003}.
DEVA is an hybrid AP$^3$M Lagrangian code, implemented with a multistep algorithm and smoothed particle hydrodynamics (SPH). 
The SPH version included in P-DEVA ensures energy and entropy conservation and, at the same time, guarantees 
a good description of the forces and angular momentum conservation. However, this advantage implies a gain in accuracy and an additional computational cost. 
Star formation (SF) is implemented through a Kennicutt--Schmidt-like law with a given density threshold, $\rho_*$, and star formation
efficiency $c_{*}$ \citep{MartinezSerrano:2008}.

The simulations have been carried out in the same periodic box of 80 Mpc side length, using $512^3$ baryonic and $512^3$ DM 
 particles. Due to computational cost, these simulations only include hydrodynamical calculation in a sub-box of 40 Mpc side. 
The evolution of matter follows the  $\Lambda$ cold dark matter ($\Lambda$CDM) model, with parameters 
$\Omega_{\rm m}=0.295$, $\Omega_{\rm b} =0.0476$, $\Omega_{\Lambda}=0.705$,
$h=0.694$, an initial power-law index $n=1$, and $\sigma_{8}=0.852$, 
taken from cosmic microwave background 
anisotropy data\footnote{http://lambda.gsfc.nasa.gov/product/map/dr3/params/
   lcdm\_sz\_lens\_run\_wmap5\_bao\_snall\_lyapost.cfm} 
\citep{Dunkley:2009}. The star
formation parameters used were a density threshold $\rho_{thres}=4.79\times10^{-25} \mathrm{g}~ 
\mathrm{cm}^{-3}$ and a star formation efficiency  $c=0.3$. 
The mass resolution is $m_{\rm bar}=2.42\times10^{7}
M_{\odot}$ and $m_{\rm DM}=1.26\times10^{8} M_{\odot}$ and a spatial resolution of $1.1$ kpc in hydrodynamical forces. 
More detailed information of these simulations can be found in \citet{Onorbe:2011}.

It is noteworthy that no explicit feedback has been implemented in these simulations, but SF regulation through the values of
the SF parameters. Nevertheless, the issues that will be discussed  
in this paper involve considerably larger characteristic scales than the ones related to stellar feedback. Therefore,  
it is unlike that the details of the star formation rate, and those of stellar feedback in particular,
could substantially alter the conclusions of this paper.

\subsection{Methods}
\label{subsec:methods}

We first describe how the LV sample around simulated galaxies has been built up. 
The first step is halo selection at $\zlow = 0.05$ 
by using  the SKID algorithm\footnote{http://www-hpcc.astro.washington.edu/tools/skid.html} 
\citep{Weinberg:1997}. 
This multi-step algorithm determines first the smoothed
density field, then it moves particles upward along the gradient of this density field using
a heuristic equation of motion that forces them to collect at local density maxima.
Afterwards, it defines the approximate group to be the set of particles identified with an
FOF algorithm with a linking length, $b$. Finally, particles not gravitationally
bound to the groups identified in the previous step are removed.

\begin{figure}
\includegraphics[width=8.4cm]{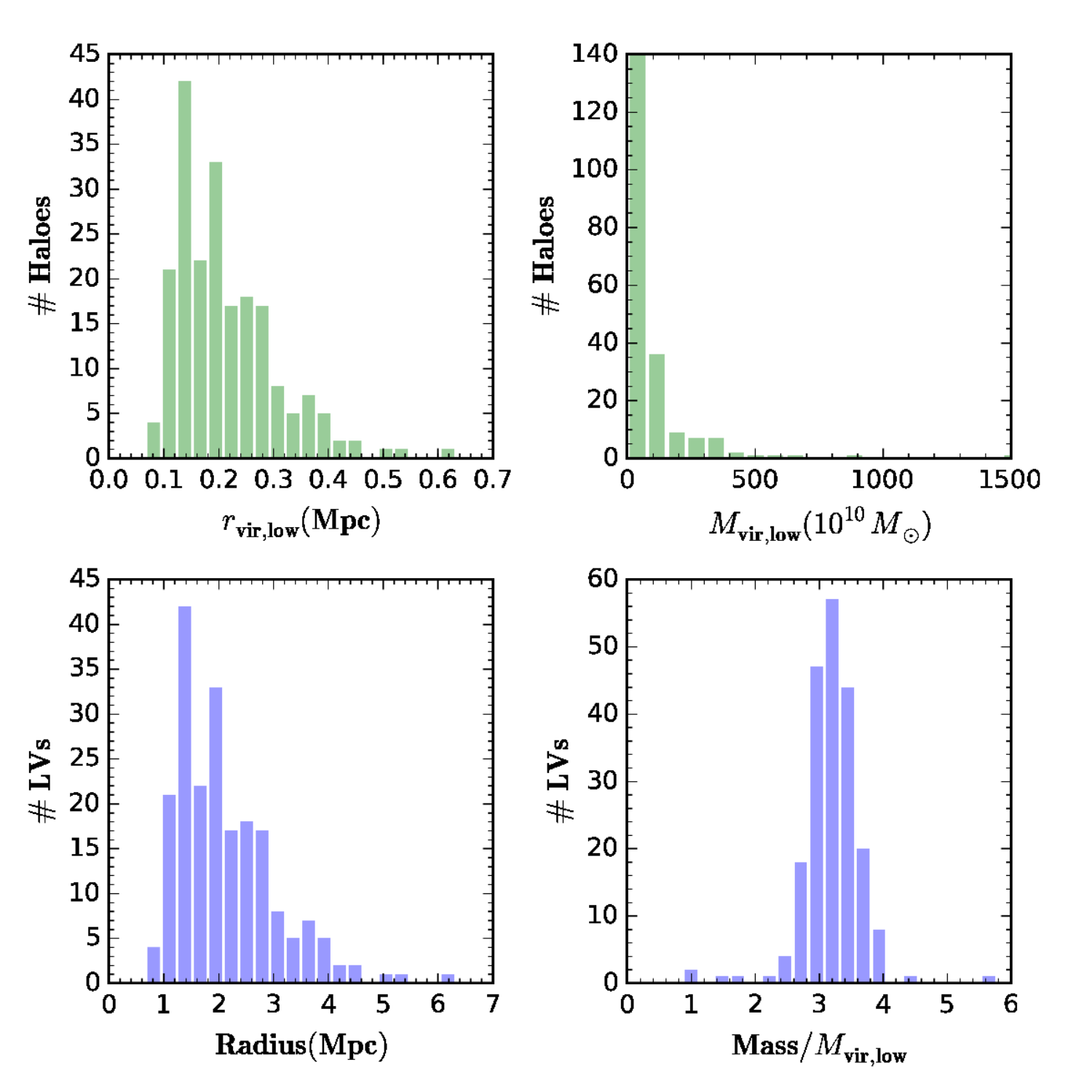}
  \caption{Upper panels show the radius and mass distribution of the galaxy haloes at $\zlow$ in our sample. Lower panels depict 
  the same information for the selected LVs. 
  }
  \label{fig:histmassrad}
\end{figure}

Specifically, we have selected a sample of 206 galaxy haloes from two runs of the GALFOBS simulations at
$\zlow $, not involved in violent events at the halo scale at $\zlow$. 
 Their virial radii $r_{\rm vir, low}$ and masses $M_{\rm vir, low}$ at this redshift 
 go from dwarf galaxies to galaxy groups, see the corresponding histograms  in Fig.~\ref{fig:histmassrad} first row. 
The virial radius ($\rvir$) is defined as the radius of the sphere enclosing an overdensity given by \citet{Bryan:1998}.

\begin{figure*}
\begin{center}$
\begin{array}{cc}
\includegraphics[width=8.8cm]{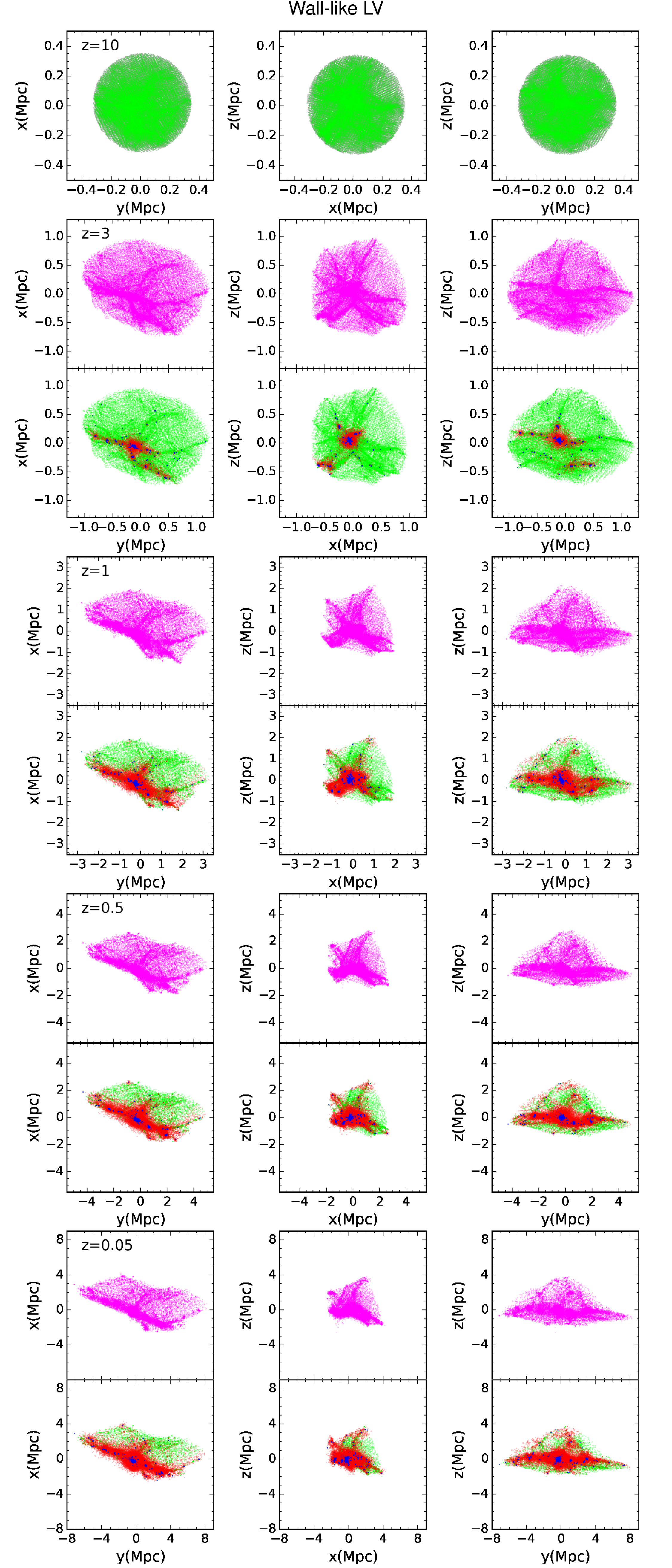} & \includegraphics[width=8.8cm]{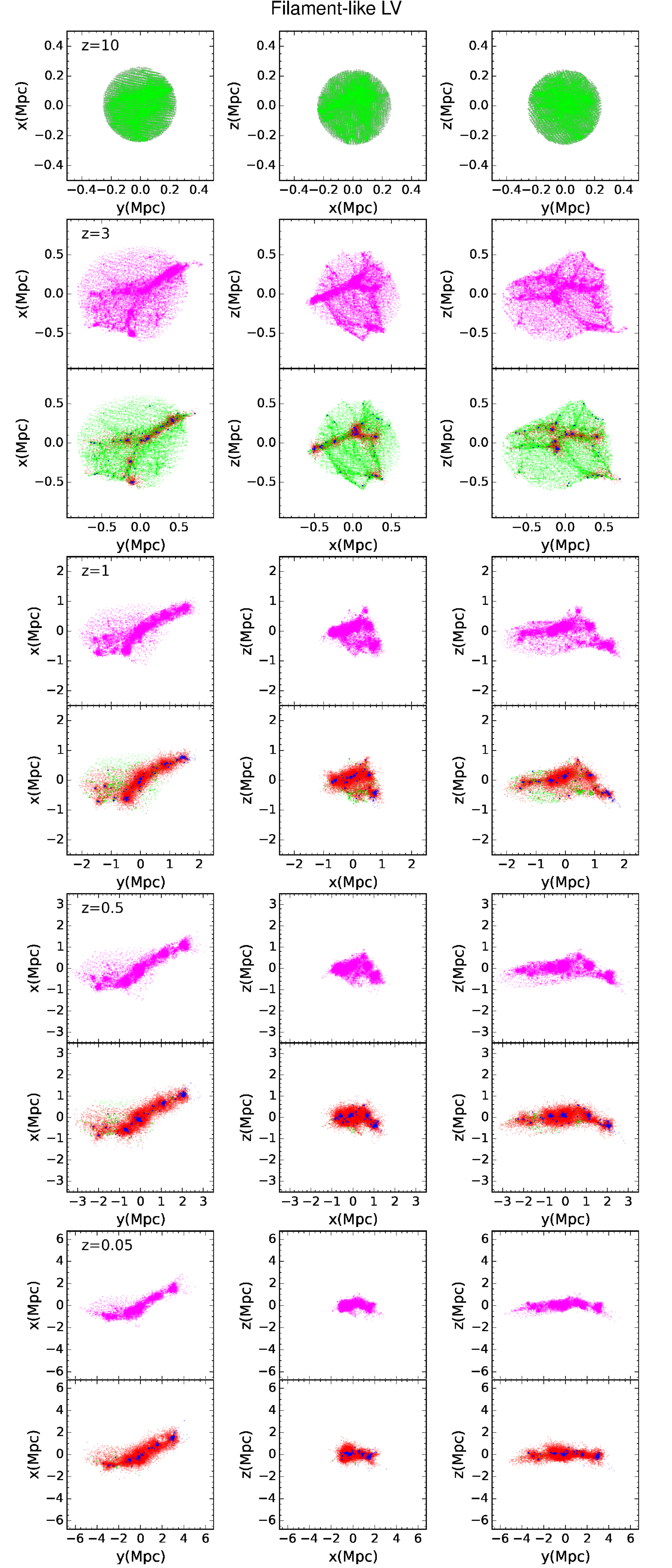}
\end{array}$
\end{center}
  \caption{Left. shape evolution of a wall-like LV from $z=10$ to $\zlow=0.05$. 
  Different columns are three  projections of the same LV, with fixed axes taken oriented along the direction of the principal axes at $\zlow$. Magenta points represent DM, green cold gas, red hot gas ($T \ge 3 \times10^4$ K) and blue stars.
  First row shows the initially spherical LV at $z=10$, where DM and cold gas are represented in the same plot.
  Second, third, fourth and fifth  group of panels illustrate the LV shape deformation across redshifts $z=3, 1, 0.5$ and $0.05$, 
 where DM and baryonic components are split in different rows. 
   Right. the same for a filament-like LV. The mass of the LVs are $8.7 \times 10^{12}M_\odot$ and $4.4 \times 10^{12}M_\odot$, respectively.
  } 
  \label{fig:lagvol}
\end{figure*}

Next, for each halo at $\zlow$ we have traced back all the particles inside the sphere defined by its respective $r_{\rm vir, low}$ 
 to $\zhigh = 10$. Using the position of these particles at $\zhigh$ we have calculated a new centre $\vec{r}_c$.
Then, we have  
selected at $\zhigh$ all the particles enclosed by a sphere of radius $R_{\rm high} = K\times r_{\rm vir, low}$, with $K = 10$  
around their respective centres $\vec{r}_c$
(see first row of Fig.~\ref{fig:lagvol}), and we have identified each of the DM and baryonic particles
within these spherical volumes. 
These particles sample the mass elements whose deformations, stretchings, foldings,
collapse and stickings we are to trace along cosmic evolution. 
They follow geodesic trajectories until they possibly get stuck and begin the formation of,
or are accreted onto, a CW structure element. 
For this reason, we have termed them  Lagrangian Volumes (LVs).
It is worth noting at this point that we are following the
evolution of individual LVs, each of them made of a fixed number of particles as they evolve. 
We do not trace the possible incorporation of off-LV mass
elements that could happen along evolution as a consequence of mergers, infalls  or other processes.  
Note also that, due to the very complex evolution of the LVs, their borders are
not well defined at $z < \zhigh$.
 Finally, a technical point to take into account is that the LVs should lie inside the hydrodynamical zoomed box.  

The choice $K=10$ is motivated as a compromise between low $K$ values, ensuring a higher number of LVs in the sample, 
and a high $K$, ensuring that LVs are  large enough to meaningfully  sample the CW emergence around forming galaxies. 
The possible effects that different $K$ values could have in our results will be discussed in Section \ref{subsec:scaleeffects}, where we conclude that $K=10$ is the best choice among the three
possibilities analysed.

Afterwards, we have followed the dynamical evolution of
these particles across different redshifts until they reach $\zlow$,
i.e., we have followed the evolution (stretchings, deformations, foldings, collapse, stickings) of a set of 206 LVs from $\zhigh$ until $\zlow$.
By construction, the mass of each of these sets of particles is constant
across evolution, and its distribution is given in Fig.~\ref{fig:histmassrad}, second row, where we also show the distribution 
of their initial sizes at $\zhigh$.    

The choice of initially {\it  spherically} distributed sets of particles aims to unveil the anisotropic nature 
 of the local cosmological evolution, illustrated in  Fig.~\ref{fig:lagvol},
 where two examples of LVs at $z=10$ and their corresponding  final shapes and orientations at $\zlow$ are displayed. 
The mass of these LVs are $8.7 \times 10^{12}M_\odot$ (left-hand panels) and $4.4 \times 10^{12}M_\odot$ 
(right-hand panels), respectively. 

In this figure we note that, in both cases, a massive galaxy appears at  $\zlow$ in the central region of the LV.
It turns out that, by construction, these galaxies are just those identified in the first step of the LV sample building-up,
 see  above. 
We also notice that the LVs have evolved into a highly irregular mass organisation, including
very dense subregions as well as other much less dense and even rarefied ones. 
 Also, some changes of orientation
of the emerging spines are visible, mainly in the lighter LV.
In addition, the initial cold gaseous
configuration at $z=10$ has been transformed into a system where stars (in blue) appear at the densest 
subregions of the LVs. Hot gas (in red) particles are also present and constitute an important
fraction of the LV mass (see $\S$\ref{FurtherEvol} for an explanation about its origin). 
We also observe that the overall LV shape on the right-hand side of Fig.~\ref{fig:lagvol} is highly elongated at $\zlow$ and has 
a prolate-like or filamentary appearance, visually spanning a linear scale  of $\sim$ 9 Mpc long by 2 Mpc  
wide, while that on the left-hand side of Fig.~\ref{fig:lagvol} still keeps a more wall-like structure. 
These shape transformations illustrate the highly anisotropic character of evolution under gravity. In this respect, it is worth mentioning that
anisotropy is a generic property of gravitational collapse
for non-isolated systems, as it was pointed out in early works by  
\citet{Lin:1965,Icke:1973} and \citet{White:1979}.

As we mentioned in $\S$\ref{Intro},  the deformation, stretching, folding, multistreaming  and collapse 
of mass elements by 
cosmological evolution is predicted and described by the ZA, while AM adds a viscosity term making
multistreaming regions to get stuck into dense configurations. 
In the following, we will introduce the mathematical methods we use to quantify the local LV transformations
illustrated in Fig.~\ref{fig:lagvol}.

To this end, we have calculated, at different redshifts,
 the reduced inertia tensor  of each LV relative to its centre of mass 
 
\begin{equation}
I_{ij}^{\rm r} =\sum_{n}m_n\frac{(\delta_{ij}r_{n}^2 - r_{i,n}r_{j,n})}{r_{n}^2}, \hspace{0.5cm} n=1, ..., N
\label{reducedI}
\end{equation}
where $r_{n}$ is the distance of the $n$-th LV particle to the LV centre of mass and $N$ is the total number of such particles. 
We have used this tensor instead of the usual one \citep{Porciani:2002a}
to minimise the effect of 
substructure in the outer part of the LV \citep{Gerhard:1983,Bailin:2005}. 
In addition, 
the reduced inertia tensor is invariant under LV mass rearrangements in radial directions relative
 to the LV centre of mass. 
This property makes the $I_{ij}^{\rm r}$ tensor particularly suited to describe anisotropic mass
deformations as those predicted by the ZA and the AM and observed in Fig.~\ref{fig:lagvol}.

 In order to measure the LV shape evolution, first, we have calculated the principal axes of  
the inertia ellipsoid, $a$, $b$, and $c$, derived from 
the eigenvalues ($\lambda_i$, with $\lambda_1 \leq \lambda_2 \leq \lambda_3$)
of the $I_{ij}^{\rm r}$ tensor, so that $a\geq b\geq c$ 
(see \citet{GonzalezGarcia:2005}),

\begin{eqnarray}
a = \sqrt{\frac{5(\lambda_2 - \lambda_1 + \lambda_3)}{2M}}, \qquad 
b = \sqrt{\frac{5(\lambda_3 - \lambda_2 + \lambda_1)}{2M}}, \\ \nonumber 
c = \sqrt{\frac{5(\lambda_1 - \lambda_3 + \lambda_2)}{2M}},
\end{eqnarray}
where $M$ is the total mass of a given LV\footnote{Note that $\lambda_1 + \lambda_2 + \lambda_3 = 2M$ and this implies $a^2+b^2+c^2=5$.}.
We denote the directions of the principal axes of inertia by $\hat{e}_i$, $i=1,2,3$, where $\hat{e}_1$ correspond to the major axis, 
$\hat{e}_2$ to the intermediate one and $\hat{e}_3$ to the minor axis.

Afterwards, to quantify the deformation of these LVs, we have computed the triaxiality parameter, $T$, \citep{Franx:1991}, defined as

\begin{equation}
T = \frac{(1-b^2/a^2)} {(1-c^2/a^2)}, 
\end{equation}
where $T=0$ corresponds to an oblate spheroid and $T=1$ to a prolate one. 
An object with axis ratio $c/a>0.9$ has a nearly spheroidal shape, while one with $c/a < 0.9$ and $T<0.3$ has an oblate 
triaxial shape. On the other hand, an object with $c/a < 0.9$ and $T>0.7$ has a prolate triaxial shape \citep{GonzalesGarcia:2009}.

We have also calculated other parameters that measure shape deformation such as, ellipticity, $e$

\begin{equation}
e=\frac{a^2-c^2}{a^2+b^2+c^2} ,
\end{equation}
that quantifies the deviation from sphericity, and prolateness, $p$

\begin{equation}
p=\frac{a^2+c^2-2b^2}{a^2+b^2+c^2}, 
\end{equation}
that compares the prolateness versus the oblateness \citep{Bardeen:1986,Porciani:2002b,Springel:2004}.
In this case, a sphere has $e=p=0$, a circular disc has $e=0.5$, $p=-0.5$ and a thin filament has $e=p=1$.
Nearly spherical objects have $e<0.2$ and $|p|<0.2$.

To sum up, we have performed the computation of the reduced inertia tensor, the principal axes of inertia, the eigendirections 
and the parameters $T, e$ and $p$ involving each of the selected LVs.
 Furthermore, we have repeated the same 
calculation for each component separately, viz. DM, cold and hot baryons. We consider hot gas as the particles  
shock heated to $3\times10^4$ K.

\section{Evolution Under the ZA or the AM}
\label{UnderEvol}

The advanced non-linear stages of gravitational instability
are described by
the {\it adhesion model} \citep{Gurbatov:1984,Gurbatov:1989,Shandarin:1989,Gurbatov:1991,Vergassola:1994}, 
an extension  of Zeldovich's
(1970) popular non-linear approximation.
In this Section, we briefly revisit them as well as some of their implications, 
useful to understand the results that will be analysed in the next sections. 

\subsection{The Zeldovich Approximation}

In comoving  coordinates, Zeldovich's approximation  is given by the so-called
{\it Lagrangian map}:

\begin{equation}
x_i(\vec q,t) = q_i + D_{+}(t) s_i(\vec q),
\label{ZAppro}
\end{equation}

where $q_i$ and $x_i,  i = 1,2,3$ are comoving  Lagrangian and Eulerian
coordinates of fluid elements or particles sampling them, respectively 
(i.e., initial positions at time $t_{in}$ and positions at
later  times $t$); $D_{+}(t)$ is the linear density growth factor.  
As already mentioned, it turns out that $s_i(\vec q)$ can be expressed as the gradient of 
the displacement potential $\Psi(\vec{q})$.

The behaviour of $D_{+}(t)$ depends on the cosmological epoch.
For the flat concordance  cosmological model (see $\S$~\ref{sec:simul}),
at high enough $z$, when the Universe evolution is suitably described by the Einstein-de Sitter model,
 $D_{+}(t) = (3/5) (t/t_i)^{2/3}$.  
Later on, when $\frac{d^{2}a}{dt^{2}} \simeq 0$ and the effects of the cosmological constant emerge 
($z_{\Lambda} \simeq 0.684$ or $t_{\Lambda}/t_{\rm U} = 0.554$ for the 
cosmological model used in the simulations analysed here), 
$D_{+}(t)$ is an exponential function of time.
Finally, when the cosmological constant dominates, we have:
\begin{equation}
D_{+}(a(t)) \propto \mathfrak{B}_x(5/6, 2/3) \left( \frac{\Omega_0}{\Omega_{\Lambda}} \right)^{1/3}\left[ 1 + \frac{\Omega_{\rm M}}{a^3 \Omega_{\Lambda}} \right]^{1/2},
\label{CurrentDmas}
\end{equation}
where  $\mathfrak{B}_x$ is the incomplete $\beta$  function, $ \Omega_0 = 1-\Omega_{\Lambda}$, $\Omega_{\rm M}$ is the 
non-relativistic contribution to $ \Omega_0$, and
\begin{equation}
x \equiv \frac{a^3 \Omega_{\Lambda}}{\Omega_0 + a^3 \Omega_{\Lambda}},
\label{xDef}
\end{equation}
describing a frozen perturbation in the limit $t \rightarrow \infty$.

Due to mass conservation, equation \ref{ZAppro} implies for the local density evolution:
\begin{equation}
\rho(\vec{r},t) = \frac{\rho_b(t)}{[1-D_{+}(t)\alpha(\vec{q})][1-D_{+}(t)\beta(\vec{q})][1-D_{+}(t)\gamma(\vec{q})]},
\label{DenZAppro}
\end{equation}
where 
$\vec{r} = a(t) \vec{x}$ is the physical coordinate,  $\rho_b(t)$ the background density, 
and $\gamma(\vec{q}) < \beta(\vec{q}) <  \alpha(\vec{q})$ are the eigenvalues of the local deformation tensor,    
$d_{i, j}(\vec{q}) = - \left(\frac{\partial s_i}{\partial q_j}\right)_{\vec{q}}$. 
Equation \ref{DenZAppro} describes caustic formation in the ZA.
Indeed, a caustic first appears when and where $D_{+}(t)\alpha(\vec{q}) = 1$ 
(i.e., a wall-like one), see details in $\S$ \ref{CWEmer}. 
Mathematically,  caustics at time $t$ can be considered as singularities in the 
{\it Lagrangian map} (see equation \ref{ZAppro} and more details in the next subsection).

\subsection{The CW Emergence in 2D}
\label{CWEmer}

The emergence of the cosmic skeleton as cosmic
evolution proceeds in the frame of the ZA is presented by \citet{Hidding:2014}.
Due to the high complexity of the formalism involved,
the authors restrict themselves to the two-dimensional equivalent of the ZA, providing us with the 
concepts, principles, language and processes needed as
a first step towards  a complete dynamical analysis of the
CW emergence in the  full 
three-dimensional space. In this subsection we give a brief summary of some of 
their results, useful to interpret some of our findings. 

In 2D, the complexity of the cosmic structure can be understood to a large extent  
from the properties of the $\alpha(\vec{q})$ landscape field,
where $\alpha(\vec{q})$ is  the largest eigenvalue of the deformation
tensor $d_{i,j}(\vec{q}), i,j=1,2$. The role
of the second eigenvalue $\beta(\vec{q})$ is  much less relevant, 
except around the places where the haloes are to form.

Of particular relevance are 
the $A_3$ lines in Lagrangian space,  because they are the progenitors of the cosmic skeleton in Eulerian space.
Geometrically they can be defined as the locus of the points where the gradient of $\alpha$ (or $\beta$)
eigenvalue is normal to its corresponding eigenvector $\vec{e}_{\alpha}$ (or $\vec{e}_{\beta}$).
Alternatively, they can also be defined as the locus of the points where $\vec{e}_{\alpha}$ (or $\vec{e}_{\beta}$)
is tangential to the contour level of the $\alpha(\vec{q})$ (or $\beta(\vec{q})$) landscape field.

The locations where collapse first occurs are around the maxima of the $\alpha(\vec{q})$
 field in Lagrangian space.
These are the so-called $A_{3}^{+}$ singularities, after Arnold's singularity classification \citep{Arnold:1983}.
They are placed on the $A_3$ lines.
Subsequently, the evolution under the ZA drives a gradual progression of Lagrangian collapsing regions, consisting, at a
given time $t$, of those points such that $\alpha(\vec{q}) = 1/D_{+}(t)$
or $\beta(\vec{q}) = 1/D_{+}(t)$, according to the 2D version of equation \ref{DenZAppro}.
 These isocontours lines are the so-called  $A_{2}^{\alpha}(t)$  and  $A_{2}^{\beta}(t)$
lines, and within them matter is multistreaming in Eulerian space, i.e., matter forms a fold caustic or pancake.

The height of the $\alpha(\vec{q})$ landscape field portrays  
the collapse time for a local mass element. Indeed, at a given time $t$, points where 
the $A_{2}^{\alpha}(t)$ and the $A_{3}^{\alpha}$ lines meet, correspond to  points
  in Eulerian
space where a cusp singularity can be found (i.e., the tip of a caustic).
The $A_{2}^{\alpha}(t)$ lines  descend on the $\alpha(\vec{q})$ landscape field as time elapses,
and in this way more and more mass elements get involved in the pancake. The pancake  grows in Eulerian space, where the
two cusp singularities at their tips move away from each other.
A similar description can be made for the $\beta(\vec{q})$ eigenvalue.
 
Note that the height of either the $A_{2}^{\alpha}(t)$ or the $A_{2}^{\beta}(t)$ lines depends only
on the $D_{+}(t)$ function, and not on the eigenvalue landscape fields.
Therefore, the higher the $\alpha(\vec{q})$ landscape field, the earlier the corresponding
pancake in the Eulerian space is formed. The same argument holds for the $\beta(\vec{q})$ eigenvalue.

Along the $A_3$ lines there are another  types of extrema.
First, we have the $A_{3}^{-}$ singularities or saddle points,
after Arnold's classification. They are in-between two $A_{3}^{+}$ singularities and are local minima
along the $A_3$ lines. They depict the places where 
two pancakes emerging from each of the $A_{3}^{+}$ points get connected, when the corresponding $A_{2}$ lines
met the $A_{3}^{-}$ singularities at their descent.
 This represents a first percolation event, and a first step towards the emergence
of the CW spine. For the aforementioned reasons, 
the higher the $\alpha(\vec{q})$ landscape field, the earlier the percolation events will occur.

The second type are the local maxima points $\vec{q}_4$, where the corresponding eigenvector is 
tangent to the $A_3$ lines, i.e., the so-called $A_4$ singularities, or swallow tail according to \citet{Arnold:1983}.
An $A_4$ singularity at $\vec{q}_4$ exists only at a unique instant $t_4$, when 
$\alpha(\vec{q}_4) = 1/D_{+}(t_4)$. At this moment, the $A_{2}^{\alpha}(t_4)$ line passes through $A_4$,
transforming the cusp singularity at the end of the Eulerian pancake into a swallow tail singularity.
After that, there are three intersections of the $A_{2}(t)$ line with two $A_{3}$ lines, giving 
three connected cusp singularities in Eulerian space. 
Therefore, the $A_{4}$ singularities are the connection points where disjoint pieces of $A_{3}$ lines
get connected in Eulerian space. Then, we get another percolation process.
Once again, as explained above, the higher the $\alpha(\vec{q})$ landscape field, the earlier
the percolation events will take place.

This short summary illustrates some aspects of
the effect that the height of the $\alpha(\vec{q})$ landscape field has
on the time when simple percolation events occur in 2D, or, in a more general scope,
when the CW spine emerges. The conclusion is simple: the higher the eigenvalue
landscape, the earlier the  percolation events take place.
A similar effect can be expected in 3D, provided that the description 
of the events connecting disjoint caustics in Eulerian space is not dramatically changed
with respect to that in 2D.

Pancake formation in Eulerian space entails an anisotropic mass rearrangement 
as matter flows normally  to the $\alpha$ (or $\beta$) pancake. These flows consist of mass elements within the
$A_{2}^{\alpha}(t)$ (or $A_{2}^{\beta}(t)$) lines in Lagrangian space, and therefore they ideally do not stop
while the $A_2$ lines keep on descending on the landscape.
Similar ideas apply to other kind of caustic formation, implying shape transformations after
the skeleton emergence.
Note that matter flows are predominantly anisotropic, except for the places where the haloes are to form, i.e. 
where flows become more isotropic.

\subsection{The Adhesion Model}
\label{AdMod}

As it is well known,
Zeldovich's approximation is not applicable beyond particle crossing,
because it predicts that caustics thicken and vanish due to
multistreaming soon after their formation.
However, $N$-body simulations of large-scale structure formation indicate
that long-lasting pancakes are indeed formed, near which particles stick, i.e 
multistreaming did not take place.
The adhesion model was
formulated to incorporate this feature to
Zeldovich's approximation,
by introducing a small diffusion
term in Zeldovich's momentum equation, in such a way that it has an
effect only when and where particle crossings are about to take place.
This can be accomplished by introducing a non-zero viscosity,
$\nu$, and  then taking the limit $\nu \rightarrow 0$.
This is  the phenomenological derivation of the  adhesion model.
 Physically motivated derivations
can be found in \citet{Buchert:1998}, \citet{Buchert:1999} and others included in
the review by  \citet{Buchert:2005}.

As in the Zeldovich approximation, in the adhesion model, 
the initial velocity field can be expressed as the gradient
of a scalar potential field, $\Phi_0(\vec q)$,
describing the spatial structure of the initial perturbation. 
It can be shown that  the solutions
for the velocity field behave just as those of Burgers' equation
\citep{Burgers:1948,Burgers:1974} in the limit $\nu \rightarrow 0$,
 whose analytical solutions are known.

The most significant characteristic of Burgers' equation solutions
  is that they are discontinuous and   hence they unavoidably develop singularities,
i.e., locations where
at a given time the velocity field becomes
discontinuous and certain particles coalesce  into {\it long-lasting}
very dense configurations with different geometries, i.e., caustics as in the ZA.
The ideas explained in $\S$~\ref{CWEmer} also apply here, but the main difference is that 
matter gets stuck forming very dense subvolumes (singularities) in Eulerian space, 
instead of forming multistreaming regions.
In this way, a singularity occurs at the time $t$
when a non-zero $d$-dimensional elemental volume $V$ around a point $\vec{q}$ 
in the initial configuration is
mapped to a $d'$-dimensional elemental volume 
around a point   $\vec{x}(\vec{q},t)$ in Eulerian space with $d'<d$. 
In a three-dimensional space, these singularities can be walls (with dimension $d'=2$),
filaments ($d'=1$) and nodes ($d'=0$).

The AM model implies that, locally, walls are the first singularities that appear, as denser small surfaces (the so-called pancakes). Later on,  
filaments form and grow  until singularity percolation and spine 
emergence 
\citep{Gurbatov:1989,Kofman:1990,Gurbatov:2012}.
The singularity pattern implies the emergence of anisotropic mass flows towards the new formed singularities.
Locally, emerging walls are the first that attract flows from voids,
then they host flows towards filaments, and, finally, filaments  are the paths of mass 
towards nodes. In this way,  at a given scale, walls and filaments 
 tend to vanish as the mass piles up at nodes.
In addition, cells associated with the deepest minima of $-\Phi_0(\vec q)$,
  swallow up  some of their neighbouring cells related to less deep minima,
 involving their constituent elements
(i.e., walls, filaments and nodes), and causing their merging, as in the ZA.
This is observed in simulations as contractive flow deformations that erase
substructure at small scales, as mentioned above,
while the CW is still forming at larger scales.

It is worth noting  that Burgers' equation solutions ensure the existence
of {\it regular points or mass elements} at any time $t$,
as those that have not yet been trapped into a caustic
at $t$.
Because of that, these regular mass elements
are among the  least dense in the density distribution.
Note, however, that due to the complex structure of the flow,
singular (i.e., already trapped into a caustic) and
regular (i.e., not yet trapped) mass elements need not be spatially segregated,
and in fact, they are mixed ideally at any scale.

\subsection{Further implications}
\label{ZAImpli} 

According to the ZA, we have 
\begin{equation}
\nabla_{\vec{q}} \cdot \vec{s} \equiv  \alpha(\vec{q}) + \beta(\vec{q}) +  \gamma(\vec{q})= \frac{5 \delta \rho}{3 \rho}(t_{in}). 
\label{LambdaPeak}
\end{equation}

As suggested by the 2D analysis made in $\S$~\ref{CWEmer}, the height of the $\alpha(\vec{q})$ landscape field
in 3D portrays
the collapse time for  local mass elements (with $\alpha(\vec{q})$ the larger $d_{i,j}$ eigenvalue at $\vec{q}$),
as well as the time when different percolation events mark the emergence of the CW spine.
Equation~\ref{LambdaPeak} indicates that the eigenvalue landscape fields are closely related to the fluctuation
 field (FF) $\frac{\delta \rho}{\rho}$ at $t_{in}$.

It is well known that the number density of the FF peaks above a given threshold is considerably
enhanced by the presence of a (positive) background field \citep{Bardeen:1986}, or, equivalently,
when a large-scale varying field is added to $\frac{\delta \rho}{\rho}$.
Equation~\ref{LambdaPeak} tells us that such background  would increase the height of the  landscape fields,
thereby speeding up percolation events responsible for the CW emergence.
Note that  denser LVs, when compared to less dense ones, can be considered as the result of adding a large-scale varying field to the
latter. Consequently, we expect that the CW elements appear and percolate   
 earlier on within denser LVs than within less dense ones.

These considerations apply to the evolution of the  $I_{ij}^{\rm r}$ eigenvectors, $\hat{e}_i(z)$, and to their
 possible dependence on mass.

Regarding shape evolution, as already emphasised,  mass anisotropically flows towards new 
singularities.
 These anisotropic mass
arrangements  make the  $I_{ij}^{\rm r}$ eigenvalues evolve. 
Thus, evolution becomes gradually extinct as anisotropic flows tend to vanish.
At small scales, the CW structure is swallowed up and
removed by contractive deformations, see previous subsection.
From a global point of view, the CW dynamic evolution somehow stops and the structure
becomes frozen as $\frac{d D_{+}(t)}{dt} \rightarrow 0$, that is after the $\Lambda$ term dominates
the expansion at $z_{\Lambda}$, see equation \ref{CurrentDmas}.
 Therefore, matter flows are expected to become on average less and less
relevant after $z_{\Lambda}$, as time elapses.

In addition, it is  expected that locally the first to vanish
are the flows associated with $\alpha(\vec{q})$, the largest eigenvalue of the local
 deformation matrix $d_{i, j}(\vec{q})$ (i.e. the flows towards walls),
 and the last to disappear are those flows
 related to $\gamma(\vec{q})$, the smallest  
 deformation matrix $d_{i, j}(\vec{q})$ eigenvalue (i.e the flows towards nodes).

Disentangling how these theoretical local predictions affect 
the global shape evolution of LVs 
demands numerical simulations. We will address these issues in the next sections.

\section{Evolution Beyond the ZA or the AM}
\label{FurtherEvol}

Some concepts, not directly described by the ZA or the AM, need to be clarified 
  in order to correctly explain Fig.~\ref{fig:lagvol} 
at a qualitative level,
as well as some results to be discussed in forthcoming sections.

\subsection{Caustic dressing}

The phenomenological Adhesion Model
tells nothing about the internal density or velocity structure of
locations where mass gets adhered.
Just to have a clue from theory, we recall that
in his derivation of a generalised adhesion-like model,
\citet{Dominguez:2000} found corrections to the momentum
equation of the ZA that regularise 
(i.e., dress) its wall singularities. These then become long-lasting structures where more mass gets stuck,
but within non-zero volumes supported by velocity dispersion coming from the energy transfer from ordered
to disordered motions.
\citep[see also][for a discussion of these effects
in terms of the viscosity, phenomenologically introduced in
the AM]{Gurbatov:1989}.
The analyses of $N$-body simulations strongly suggest that
any kind of flow singularity gets dressed \citep[i.e., not only at pancakes,
as it has been analytically proven by][]{Dominguez:2000}.  

\subsection{Gas in the cosmic web}
\label{GasCW} 

When gas is added, the energy transfer from ordered
to disordered motions around singular structures includes
the transformation of velocity dispersion 
into internal gas energy (heating) and pressure. Then,   
energy is lost through gas cooling, mainly at the densest pieces of the CW, making them even denser. 
However, as already said in $\S$\ref{AdMod}, singular (i.e., dense) and regular 
(i.e., not yet involved in singularities, low density) mass elements are mixed at any scale.
Therefore, low-density gas is heated too, and, in addition, pressurised. 
The consequences of these processes cannot be deciphered from theory, but previous analyses 
of cosmological hydrodynamical simulations in terms of the CW \citep[see, for example][]{DT:2011}
suggest that dressing acts on any kind of flow singularity, 
i.e., also on filaments and nodes. Moreover, these authors conclude that, at (node-like) halo collapse,
cooling of low-density gas is so slow that most gravitationally  heated
 gas is kept hot until  $z =0$.
In any case, because hot gas is pressurised, no anisotropic mass inflows towards singularities
  can be expected within the hot gas component, on the contrary, possible anisotropic,
pressure-induced hot gas outflows  are expected from them. 
These expectations will be explored in the following sections.  

On the other hand, at the densest gas locations, cold gas is transformed into stars with an efficiency $\epsilon$
when the density is higher than a threshold. 
In this way, the hot gas component and the stars,  observed in Fig.~\ref{fig:lagvol}, arise.

\subsection{A visual impression of LV evolution}

Fig.~\ref{fig:lagvol} gives us a first visual impression of the evolution of the initially spherical LVs.
The former considerations above make it easier a qualitative interpretation of what these figures 
show. Indeed, the gradual emergence of a local skeleton stands out in both of them, including web-element
mergings and some rotations too. Finally, 
at $\zlow$, we see an elongated structure, either in the DM, cold or hot baryonic components,
 where different spherical configurations appear, with a stellar component at the centre of most of 
them\footnote{We note that there is a component effect, namely different components (i.e. DM, cold and hot baryons)  evolve dissimilarly.}.
A high fraction of hot gas component (but not its whole mass) is related to these spheres. 
This complicated structure comes from wall and filament formation, according to the AM, and its  dressing 
and eventual fragmentation into clumps. Clumps are in their turn dressed. 
Note also that, at each $z$, a fraction
of the matter is not yet involved into singularities. 
Therefore, evolution leads to:  
(i) a DM component  sharing both a diffuse and a dressed singularity configuration, with the particularity that the 
LV diffuse component present at redshift $z$ has not yet been involved in any singularity at $z$, 
(ii) a complex cold gas component, sharing also  a diffuse as well as a dressed singularity configuration, 
but with a more concentrated distribution than that of the DM, because
gas can lose energy by radiation and
(iii) a complex hot gas distribution. As explained in $\S$~\ref{GasCW}, diffuse gas is gravitationally heated at collapse events, 
but, as will be shown in $\S$~\ref{sec:CompEff}, it is 
not involved in important anisotropic mass rearrangements.

To further advance,  we need a quantitative  analysis of LV evolution. This is the subject
of the next sections.

\section{Anisotropic Evolution: Eigenvectors of the mass distribution}
\label{EigenEvol}

According to the AM, mass elements are anisotropically deformed and a fraction of them pass through
one or several singularities in sticking regions. For each mass element placed at a 
Lagrangian point $\vec{q}$, accretion at high $z$ preferentially
occurs along the eigenvector corresponding to the largest eigenvalue of the symmetric deformation matrix at $\vec{q}$,
$d_{i, j}(\vec{q}) = - \left(\frac{\partial s_i}{\partial q_j}\right)_{\vec{q}}$.

\begin{figure}
\includegraphics[width=8cm]{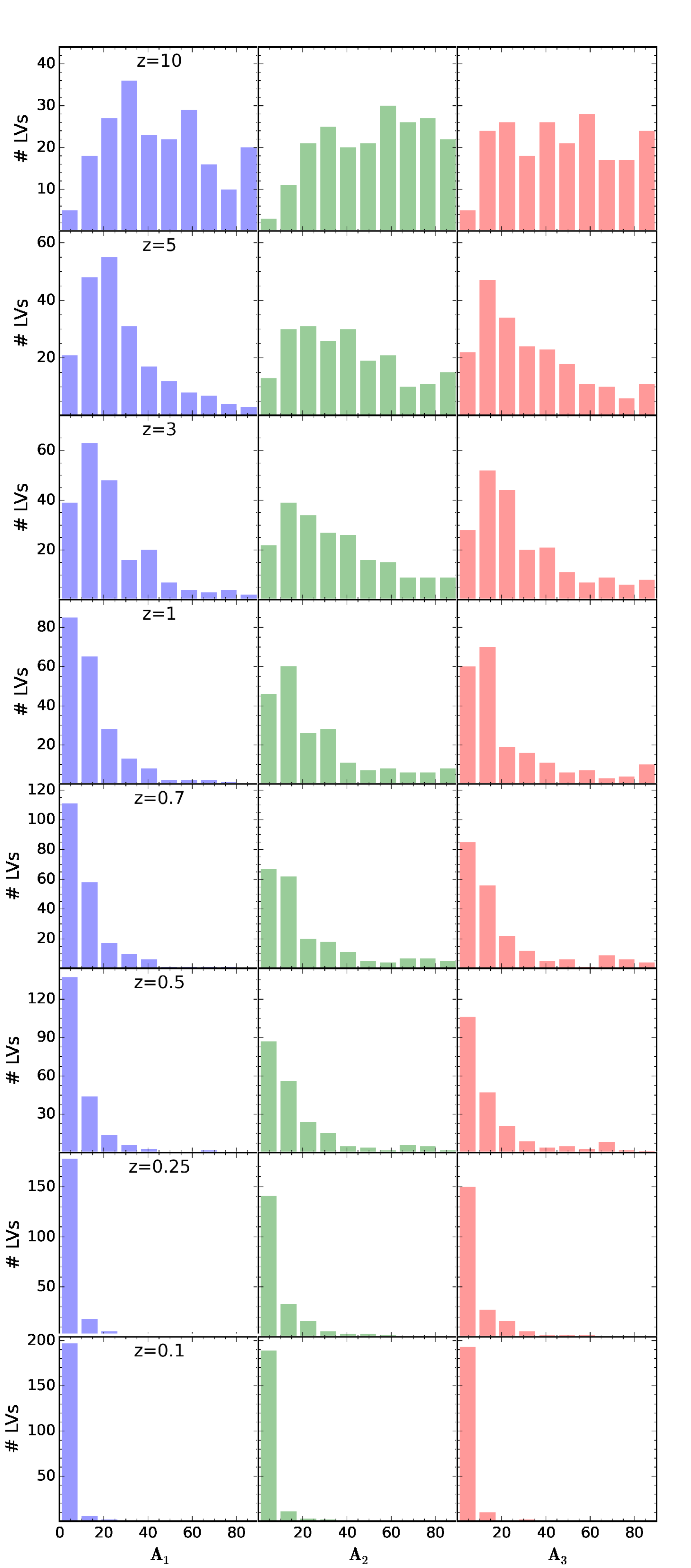}
  \caption{Evolution across redshifts of the $A_i$ distribution, where $A_i$ is the angle formed by the eigenvectors
 $ \hat{e}_i^{\rm tot}(z)$ and $\hat{e}_i^{\rm tot}(z_{\rm low})$, with $i=1,2,3$, and where `$\rm tot$' stands for the eigenvectors of
 the $I_{ij}^{\rm r}$, calculated with all the LV components. 
}
  \label{fig:direc-evol}
\end{figure}

\begin{figure}
\includegraphics[width=8.4cm]{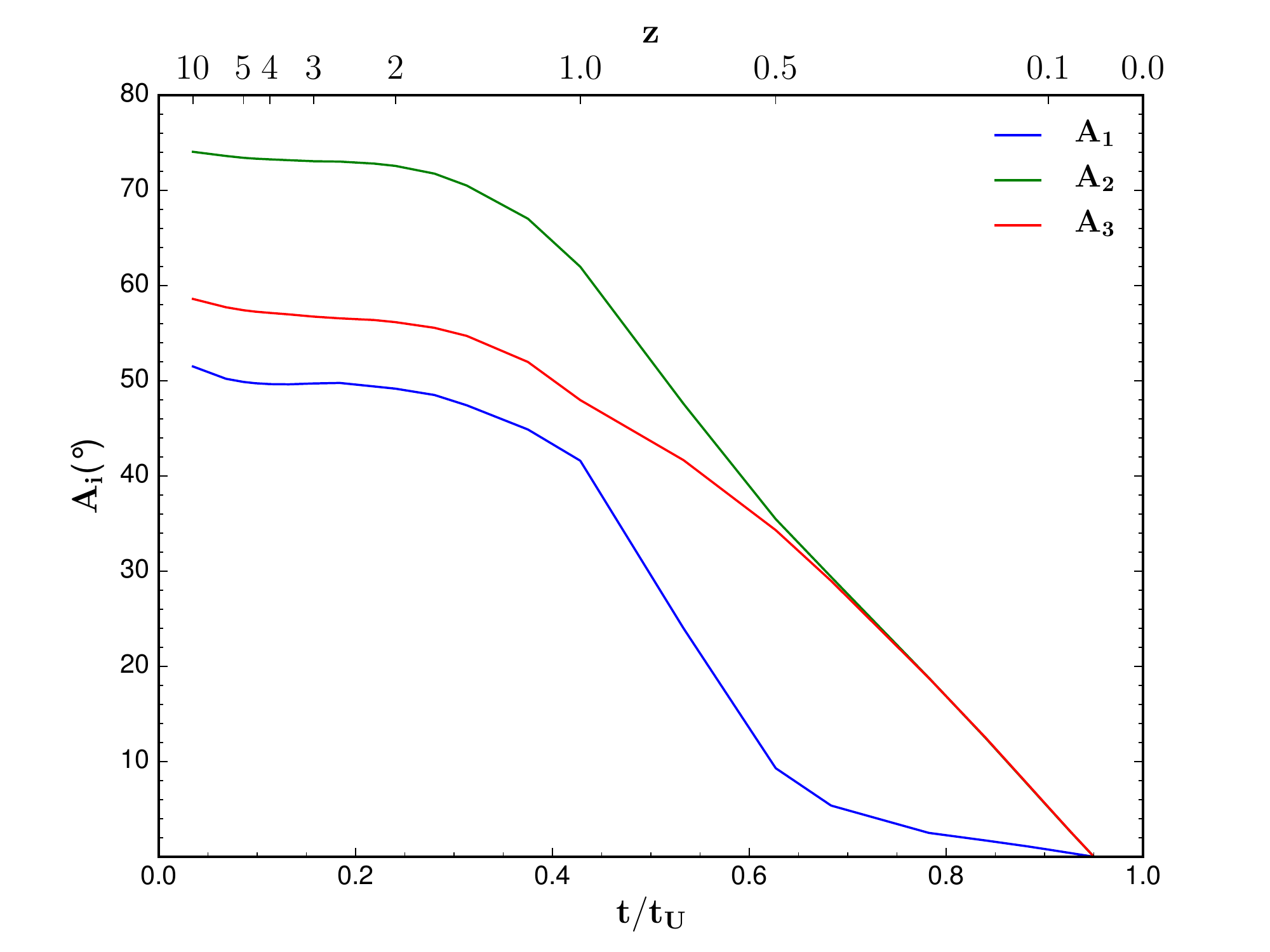}
\caption{An example of the $A_i(t)$ evolution, where $A_i$ is 
the angle formed by the eigenvectors
 $ \hat{e}_i^{\rm tot}(z)$ and $\hat{e}_i^{\rm tot}(z_{\rm low})$, 
with $i=1, 2, 3$ and $t$ is given in terms of the age of the Universe ($t_{\rm U}$).
}
\label{fig:angAi}
\end{figure}

Taking the LV as a whole, the $I_{ij}^{\rm r}$ eigenvector $\hat{e}_3^{\rm tot}(z)$ which corresponds to its larger eigenvalue,
 $\lambda_3(z)$ at a given redshift $z$, defines the direction along which the overall LV elongation has been maximum
until this $z$. Similarly,
$\hat{e}_1^{\rm tot}(z)$ corresponds to the direction of overall minimum stretching of the LV up to a given $z$.
It is very interesting to analyse whether or not there exists a change in such directions as cosmic evolution proceeds.
In Fig.~\ref{fig:direc-evol}, we show the histograms for the quantities
$A_i(z)$, the angle formed by the eigenvectors
 $ \hat{e}_i^{\rm tot}(z)$ and $\hat{e}_i^{\rm tot}(z_{\rm low})$, with $i=1,2,3$, where `$\rm tot$' stands for the eigenvectors of
the $I_{ij}^{\rm r}$ tensor corresponding to the total mass of the LV, at redshifts $z=10,5,3,1,0.7,0.5,0.25,0.25,0.1$.
That is, we measure the deviations from the eigendirections at a given $z$ with respect to the final 
ones\footnote{Note that only two out of the three $A_i$ angles are independent in such a way that if for instance $A_1=0$ then $A_2=A_3$.}.
We see that on average these directions are frozen at $z_{\rm froz} \sim 0.5$, in such a way that
only a few LVs change the eigenvectors of their total mass distribution at 
$z \le z_{\rm froz}$, while at $z \ge z_{\rm froz}$ more and more LVs do it.
This behaviour is illustrated by Fig.~\ref{fig:angAi}, where the evolution of the $A_i(t)$ for a typical LV case is plotted.
We observe that $A_i(z)$ smoothly and gradually vanish before $t/t_{\rm U} = 1$, this behaviour being common
to all the LVs.
 
This is particularly interesting, because as we will see in Figs
\ref{fig:prinaxes} and  \ref{fig:axisratios} 
the evolution of the $I_{ij}^{\rm r}$ eigenvalues (or, equivalently, that of its
principal axes of inertia $a, b, c$), also declines before  $t/t_{\rm U} = 1$.

\begin{figure}
\includegraphics[width=8.7cm]{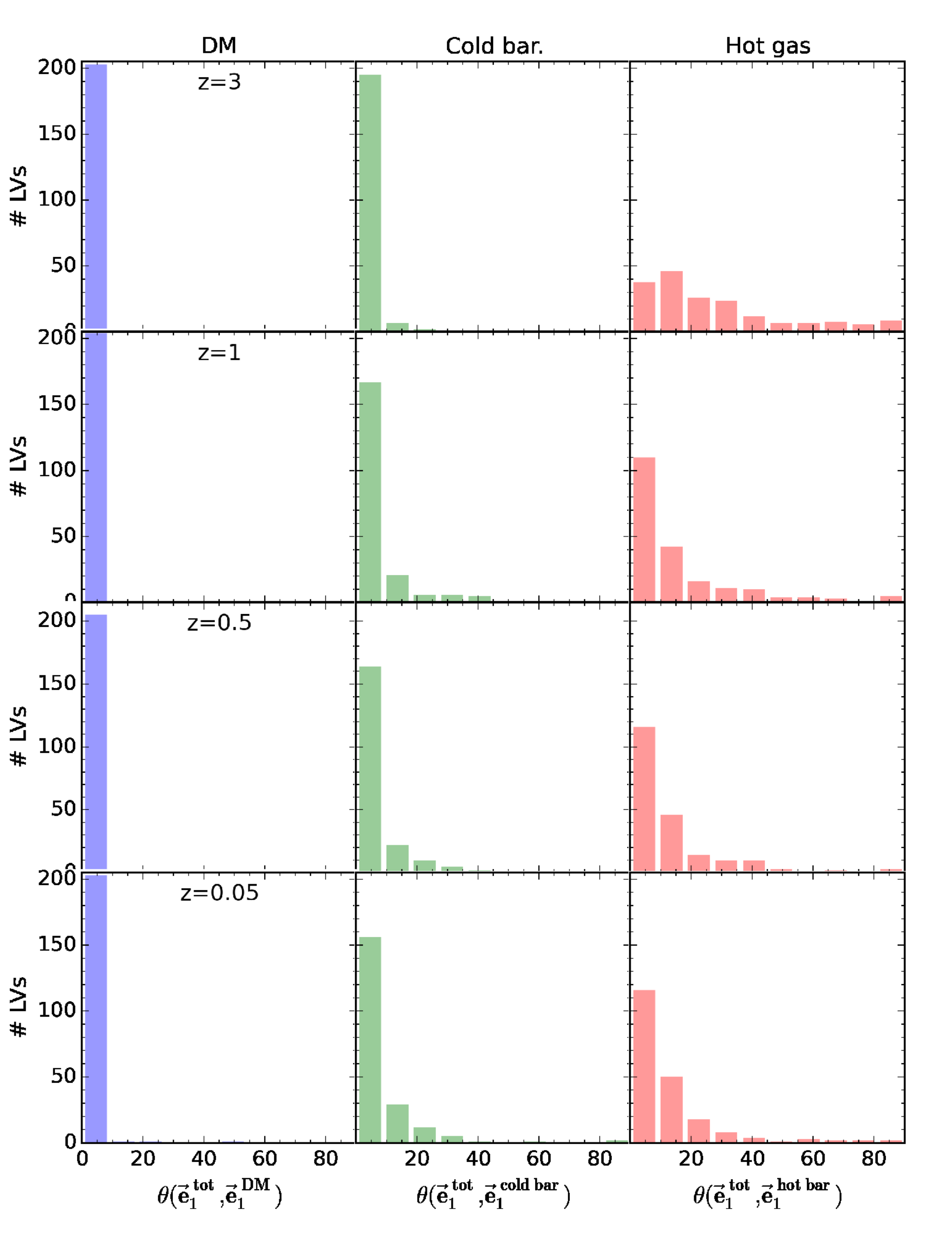}
  \caption{Distributions of the angles formed,  at several redshifts, by the direction of the $ \hat{e}_1^{\rm tot}(z)$
 axis of inertia that arise from the overall
 matter distribution with the same axis calculated with the different components.
}
\label{fig:histangei}
\end{figure}

It is also important to investigate if there exists a component effect in the freezing-out of the eigendirections.
With this purpose, we have compared the directions of the principal axes of inertia that arise from the whole mass distribution with the ones 
derived from every component at different redshifts (see Fig.~\ref{fig:histangei}). We have
found that the latter are mainly parallel to $\hat{e}_i^{\rm tot}$ in the DM and cold baryon cases.
Concerning hot gas, the distribution of the angles, $\theta_i$, formed by $\hat{e}_i^{\rm tot}$ and $\hat{e}_i^{\rm hot~bar}$, the eigenvectors 
of the hot gaseous component, starts nearly uniform and as time elapses a peak around $0$\textdegree~arises, 
as we can observe in Fig.~\ref{fig:histangei} for the $\hat{e}_1$ case.

This means that DM dynamical evolution determines
the preferred directions of LV stretching, and cold gas particles
closely follow them.  Hot gas particles (in this case, as explained in $\S$\ref{FurtherEvol}, gaseous particles
not trapped into singularities and heated by gravitational collapse), on the contrary, do not follow DM
evolution at high redshifts, but they trace at any $z$ the locations where 
mass sticking  events have taken place. Indeed,
as explained in $\S$\ref{FurtherEvol}, gas gravitational heating is due to the transformation of the ordered flow energy into internal energy
at CW element formation.

\section[]{Anisotropic evolution: Shapes}
\label{sec:results}

Before we focus on the statistical analysis of our results, we present the shape evolution of some selected LVs 
in order to show how they acquire their filamentary or wall shape. 
Then, we analyse the shape evolution of all the objects in our sample, by considering component as well as mass effects.
 To that end, LVs are grouped according to 
their mass, $M$, into three bins, massive ($M\geq5\times10^{12} M_\odot$), intermediate mass 
($5\times10^{11}\leq M<5\times10^{12} M_\odot$) and low-mass LVs ($M<5\times10^{11} M_\odot$). 

\subsection{Two particular examples of shape evolution}
\label{Shape-examples}

\begin{figure}
\includegraphics[width=8.5cm]{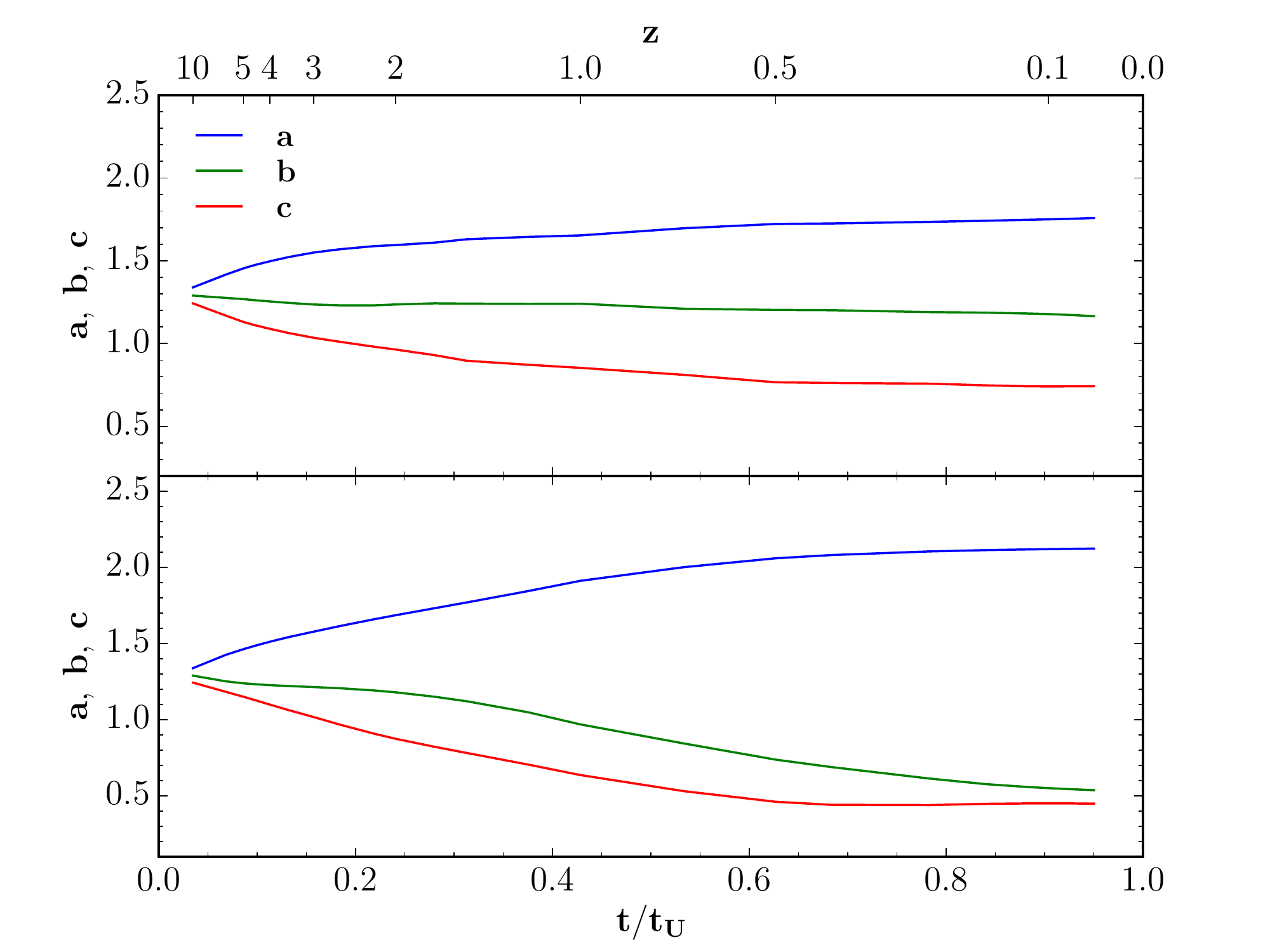} 
  \caption{Evolution of the principal axes of inertia for two LVs. 
  Top, LV on the left-hand side of Fig.~\ref{fig:lagvol}, with a wall-like structure. 
  Bottom, LV on the right-hand side of Fig.~\ref{fig:lagvol}, which acquires a filamentary shape.
 } 
\label{fig:prinaxes}
\end{figure}

\begin{figure}
\includegraphics[width=8.5cm]{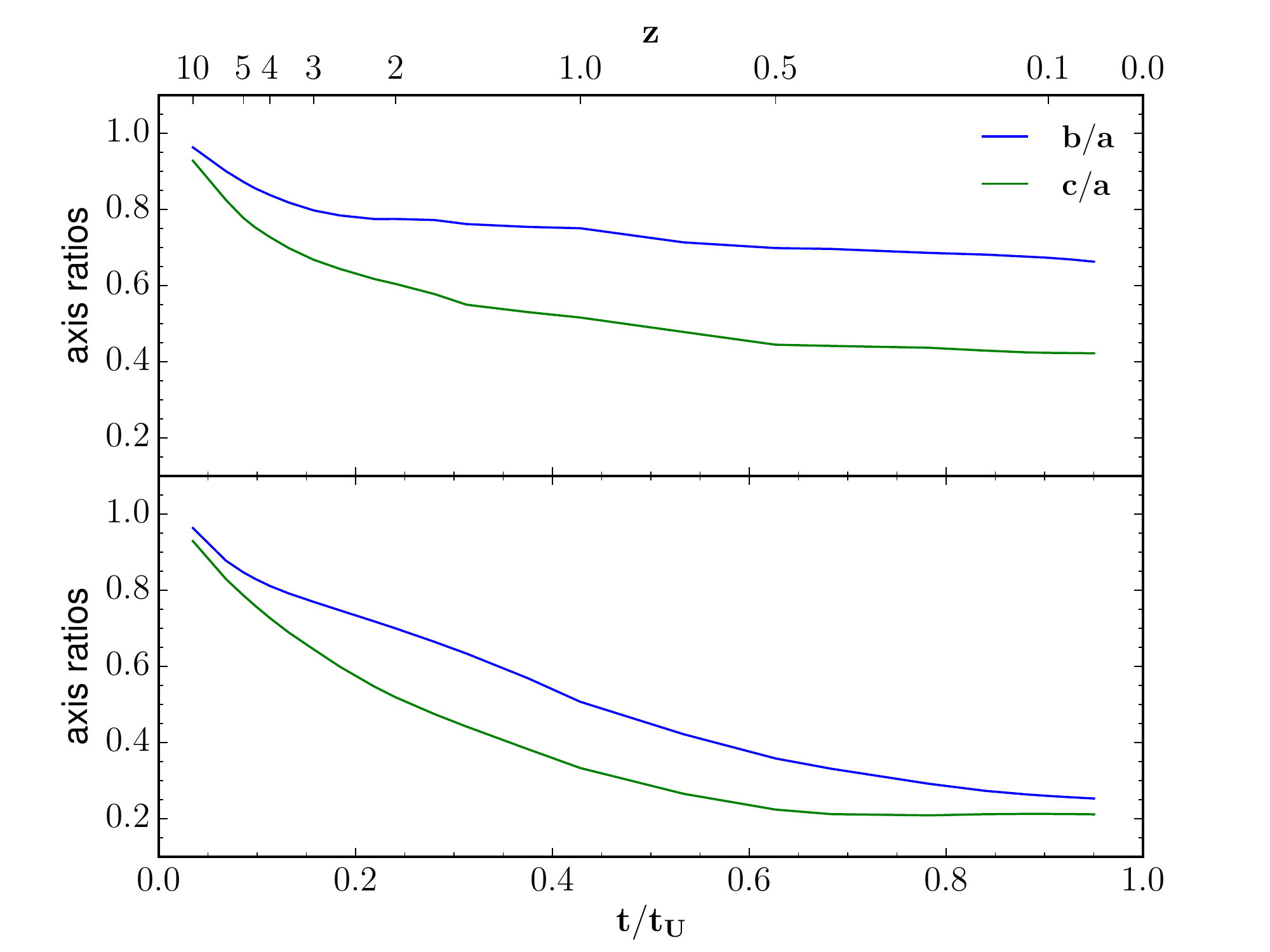}
  \caption{Axis ratio evolution of the Lagrangian volumes of Fig.~\ref{fig:prinaxes}. The upper plot shows the evolution towards an 
  oblate shape and the lower plot shows an LV that acquires a prolate shape.
  } 
  \label{fig:axisratios}
\end{figure}

In Fig.~\ref{fig:prinaxes}, we exemplify the evolution of the principal axes of the inertia ellipsoid for the LVs   
of Fig.~\ref{fig:lagvol}.  
The upper plot (LV on the left-hand side of Fig.~\ref{fig:lagvol}) illustrates an LV that has two axes that expand across time, 
i.e., it has a flat structure. 
The lower plot corresponds to the LV on the right-hand side of Figure~\ref{fig:lagvol} and portrays the
case in which the major axis grows while the other two axes are compressed, giving in consequence a prolate shape. 
This result can 
also be inferred from Fig.~\ref{fig:axisratios}, where we can see the evolution of the axis ratios $b/a$ and $c/a$ 
for the same LVs of Fig.~\ref{fig:prinaxes}. In the lower plot of Fig.~\ref{fig:axisratios}, we observe that the two minor axes end up close to each other in 
length, therefore the LV has a filamentary structure. The upper plot, in contrast, has the minor axis significantly shorter than the 
other two, hence having an oblate shape.

A remarkable result is the continuity of the $a(t), b(t)$ and $c(t)$ functions for all the LVs, 
with no mutual exchange of their 
respective eigendirections across evolution, i.e., the local skeleton 
is continuously built up, in consistency with \citet{Hidding:2014}.

\subsection{Generic trends of shape evolution}
\label{Shape-Evol}

\begin{figure*}
\includegraphics[width=16.1cm]{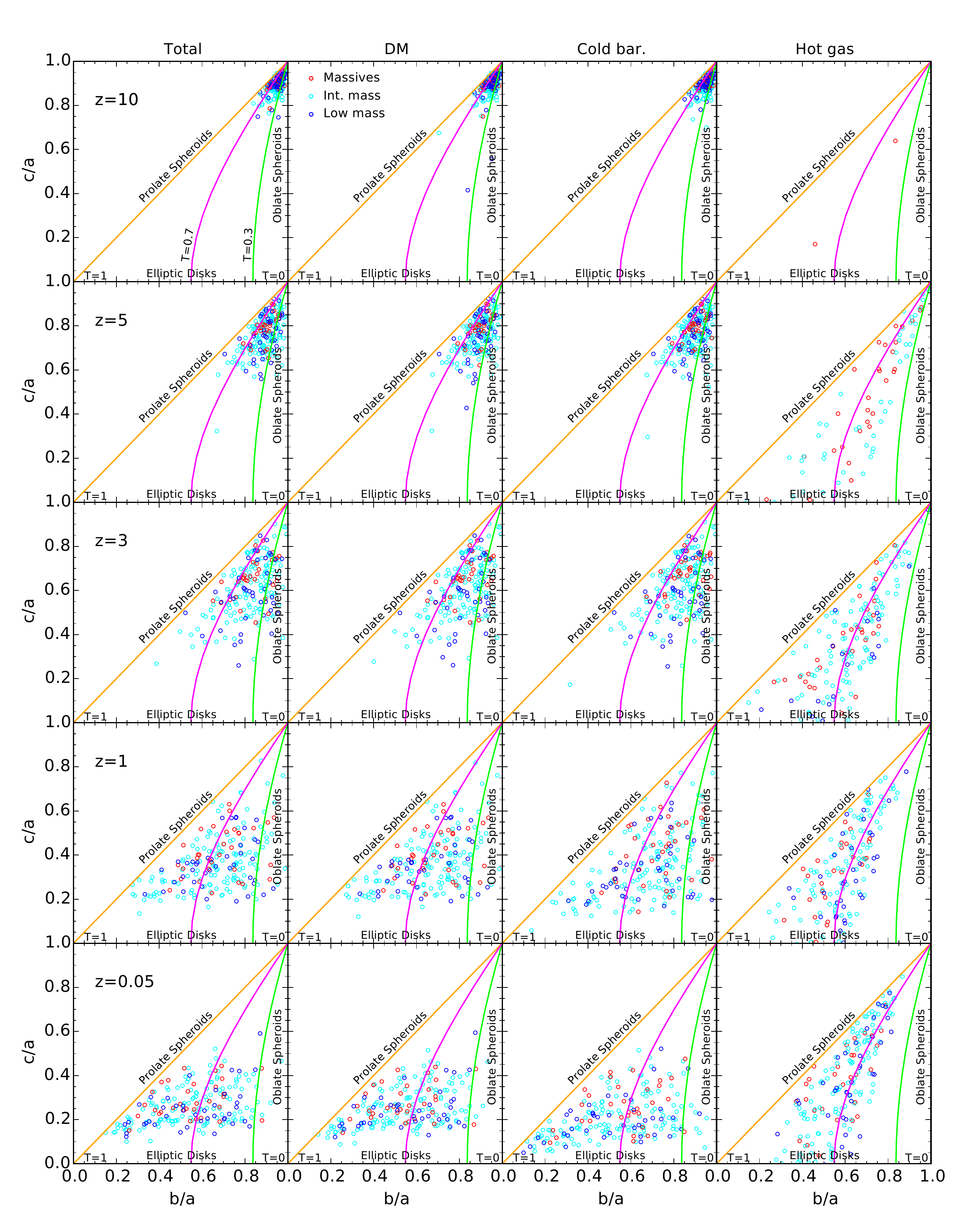}
  \caption{Axis ratio evolution of all the selected LVs, where coloured circles indicate different mass range.
  Massive LVs with 
   $M\geq5\times10^{12} M_\odot$ are represented in red, LVs with intermediate mass, $5\times10^{11}\leq M<5\times10^{12} M_\odot$, 
   in cyan and low-mass LVs, $M<5\times10^{11} M_\odot$, in blue. The orange line correspond to $T=1$, i.e., to a prolate spheroidal 
   shape. Objects with $c/a<0.9$ and $T>0.7$ (magenta line) have a prolate triaxial shape and LVs with $c/a<0.9$ and 
   $T<0.3$ (green line) are prolate triaxial ellipsoids. We show the axis ratios obtained with the total number of particles, the 
   axis ratios of DM particles, and the axis ratios found for cold and hot baryons.
}
  \label{fig:axisratiosevol}
\end{figure*}

In this subsection, the generic trends of shape evolution are examined at a qualitative level. 
In Fig.~\ref{fig:axisratiosevol}, where the axis ratios are plotted,  
we can note that the selected LVs are gathered on the nearly spherical zone ($c/a\geq 0.8$) 
by construction, except the hot gaseous component. As time elapses, LVs are deformed, and their evolution is shown as 
they move down inside the triangle described by the axes $b/a$, $c/a$ and $T=1$ (orange line). Accordingly, at $z=0.05$ they tend to be spread over the 
triangle. 
Note that intermediate mass and low-mass objects evolve faster than the massive ones. At $\zlow$, DM is preferentially 
located in the $T>0.3$ and $c/a<0.4$ region, therefore we end up with more prolate systems than oblate objects. This assertion is valid for 
the total, DM and cold baryons axis ratio evolution. 
In contrast, hot gas does not seem to show a remarkable evolution effect as it appears   
populating roughly the same regions of the aforementioned triangle at redshifts $10, 5$ and $3$, 
and later on, excluding either the  oblate area on the right or the prolate one at the left bottom corner of the triangle.

\begin{figure*}
\includegraphics[width=16.1cm]{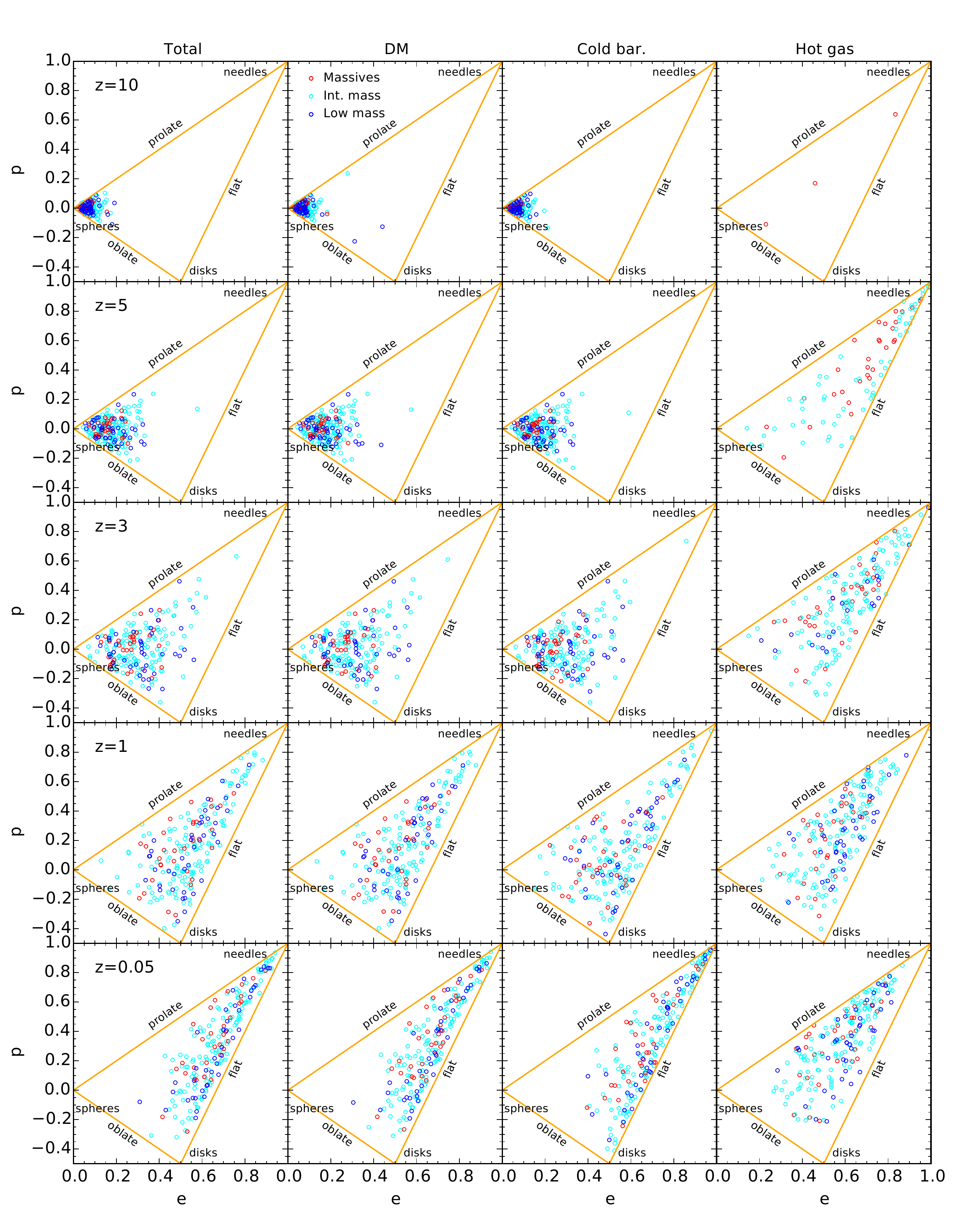}
\caption{Prolateness-ellipticity plane for the reduced inertia tensor of the selected LVs for redshifts $10, 5, 3, 1$ and $0.05$.  
Massive LVs with $M\geq5\times10^{12} M_\odot$ are represented in red, LVs with intermediate mass, 
$5\times10^{11}\leq M<5\times10^{12} M_\odot$, in cyan and low-mass LVs, $M<5\times10^{11} M_\odot$, in blue. 
The orange lines correspond to ultimate shapes, $e=p$ (prolate spheroids), $p=-e$ (oblate spheroids) and $p=3e-1$ (flat objects).
}
\label{fig:prolatellip}
\end{figure*}

The shape evolution of the LV mass distribution is also shown in Fig.~\ref{fig:prolatellip}, where shape 
distortions are represented in the prolateness-ellipticity plane. In this case, LVs move inside the triangle bound by the lines, 
$e=p$ (prolate spheroids), $p=-e$ (oblate spheroids) and $p=3e-1$ (flat objects). We observe the same pattern as in 
Fig.~\ref{fig:axisratiosevol}, for the total components, DM and cold baryons. In other words, 
initially spherical systems, concentrated on one corner of the triangle, evolve across redshifts 
filling up the triangle, so that, at $z=0.05$, we end up with a high percentage of prolate triaxial objects, $\sim 83\%$ for the total 
inertia ellipsoid. We have also found that $\sim 91\%$ of the selected LVs have extreme total ellipticities ($e>0.5$), 
while only $8\%$ have moderate ones. A significant percentage of the analysed objects are extremely prolate, $\sim 31\%$, 
that is, they have a thin filament-like shape. At $z=0.05$, we can find systems close to the flat limit, specially in the case of 
cold baryons. 
As in the previous figure, hot gas does not present a remarkable 
evolution effect after $z=1$. At higher $z$s, however, the hot gas in some LVs
show needle-like as well as flat shapes (see panels corresponding to $z=5$ and, to a lesser extent, at $z=3$), 
but these shapes do not appear anymore at lower $z$s.

Figs~\ref{fig:axisratiosevol} and~\ref{fig:prolatellip} nicely show generic trends of shape evolution.
More elaborated, quantitative analyses of component and mass effects are given in the next sub-sections.

\subsection{Component effects}
\label{sec:CompEff}

\begin{figure*}
\includegraphics[width=13cm]{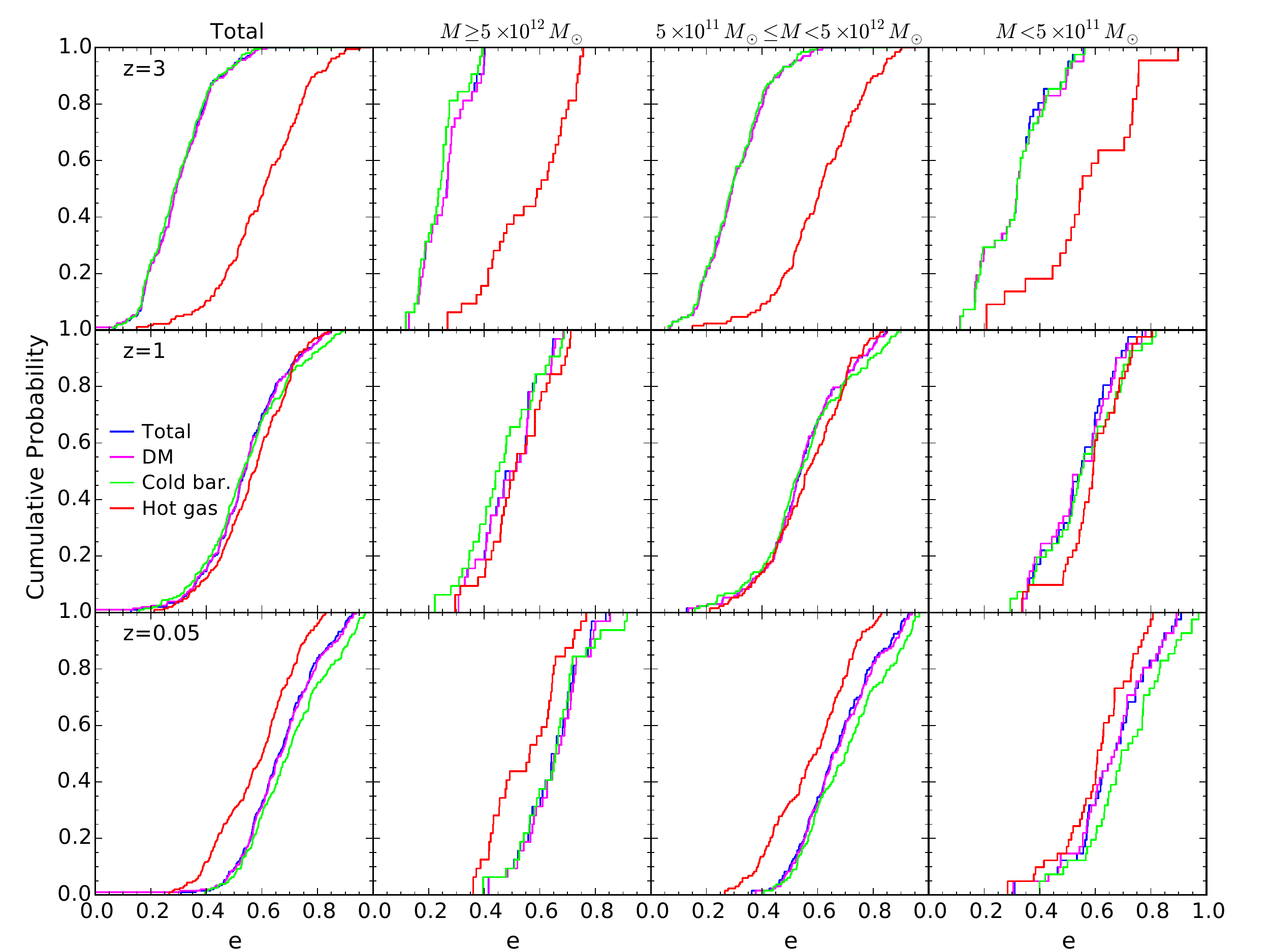}
\caption{Cumulative distribution function of the ellipticity parameter, $e$, portraying component effects and their evolution in different mass bins.  
Each column shows the distribution binned according to 
the LV mass. Plots in the first column are calculated for the total number of LVs. Rows represent different redshifts. 
The code colour used in each plot is as follows, results obtained with the total reduced inertia tensor are presented in blue, 
DM results in magenta, cold baryons in green and hot gas in red.}
\label{fig:cumhistecomp}
\end{figure*}

\begin{figure*}
\includegraphics[width=13cm]{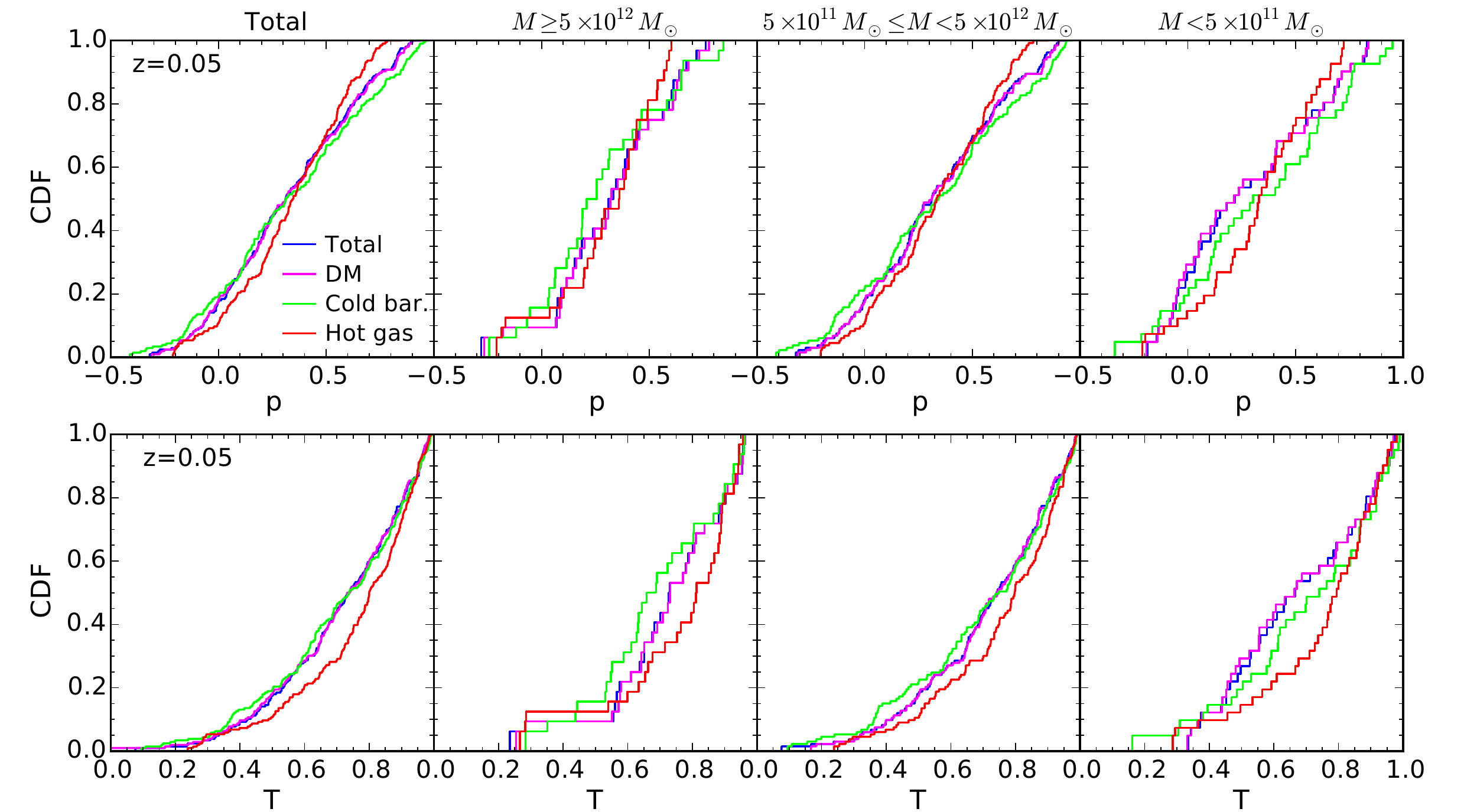}
\caption{Upper panels, CDF of the prolateness parameter, $p$, at $\zlow$. Lower panels, CDF of the triaxiality parameter, $T$, parameter at $\zlow$.
Each column shows the distribution binned according to 
the LV mass. Plots in the first column are calculated for the total number of LVs.  
The code colour is as in Fig.~\ref{fig:cumhistecomp}.}
\label{fig:cumhistpTcomp}
\end{figure*}

In order to quantitatively determine if there is a component effect  on the LV shape evolution 
(i.e., whether DM, hot and cold baryons behave dissimilarly), we represent the cumulative distribution 
function (CDF) of the $e, p$ and $T$ parameters in Figs.~\ref{fig:cumhistecomp} and \ref{fig:cumhistpTcomp}.

Each row in Fig.~\ref{fig:cumhistecomp} shows the cumulative probability  
of the $e$ parameter calculated for DM, cold baryons, hot gas and the total components at a given redshift. The first column 
depicts the result obtained for all the LVs and the other columns display our findings split according to the binning in LV mass.
As we can observe, the DM and cold baryonic components move from low ellipticities or high sphericities at high redshifts towards higher ellipticities at $\zlow$. 
As a result, these components acquire a filament-like structure (see Fig.~\ref{fig:cumhistecomp}). 
Note that cold baryons and 
DM exhibit approximately the same behaviour as time elapses. 
At $\zlow$ cold baryons are slightly more prolate than the DM component, specially in the case of low-mass LVs.
On the other hand, the hot gaseous component does almost not experience an evolution effect, as can be noted from the ellipticity CDFs 
in Fig.~\ref{fig:cumhistecomp}, whether or not we group the LVs according to their mass.
Hot gas has an $\bar{e}\sim 0.57$ since $z=2$, and does not present any preference for either a spherical or a filamentary structure.

Similar conclusions can be extracted from the DM, cold baryon and hot gas prolateness CDFs 
(see first row of Fig.~\ref{fig:cumhistpTcomp}). In this case, hot gas has an  
$\bar{p}$ ranging from $0.25 - 0.34$ since $z=2$. An important difference with respect to the ellipticity CDFs is that at $\zlow$, 
hot gas cumulative probabilities show a small deviation from cold baryons CDFs which is bigger in the low-mass bin, while in the $e$ case these 
components exhibit a large deviation from each other.

Triaxiality CDFs show a tendency of cold baryons to have a prolate shape independently of the 
mass binning at $z=3$. We observe the same displacement of DM and cold baryon CDFs across redshifts, previously noted 
from ellipticity and prolateness cumulative probabilities. 
Concerning hot gas, it has an $\bar{T}$ in the range $0.69 - 0.76$ since 
$z=2$, showing almost no changes thereafter. 
This displacement causes that the difference between DM, cold and hot baryons CDFs appears greatly diminished at $z=1$. 
This fact can also be noticed from ellipticity and prolateness CDFs. 
It is noteworthy that the cold baryon triaxiality cumulative probability of the massive LV bin is delayed with respect to the DM CDF at $z=1$. This 
difference is kept at $z=0.05$ (see lower panels of Fig.~\ref{fig:cumhistpTcomp}), this is also true for the prolateness case. 
On the contrary, at $\zlow$ DM CDF appears delayed with 
respect to cold baryons for the low-mass bin.

\subsection{Mass effects}

\begin{figure*}
\includegraphics[width=13cm]{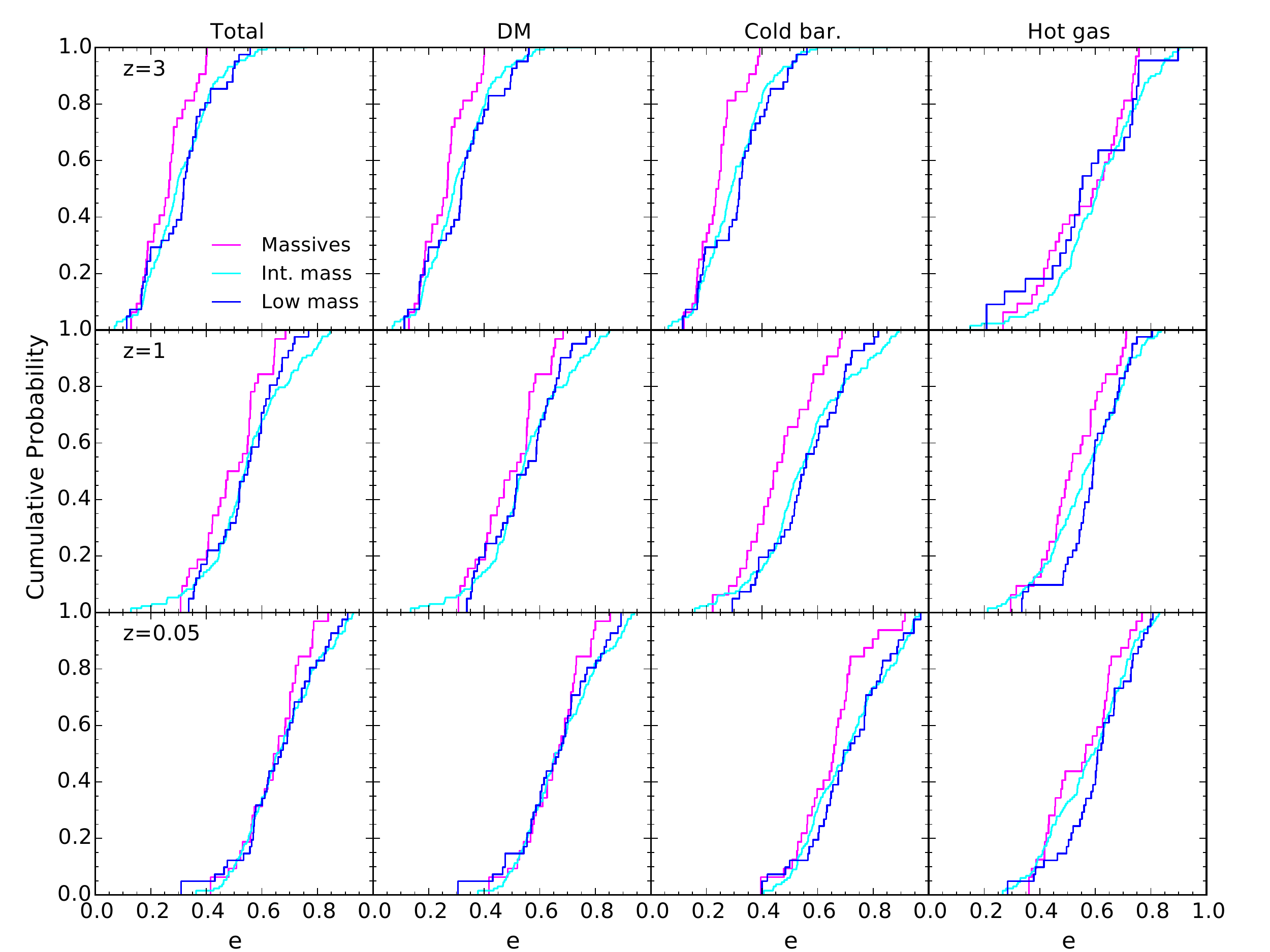}
\caption{Cumulative distribution function of the ellipticity parameter, $e$, illustrating mass effects and their evolution according to the LV components. 
Each column displays the distribution binned according to the 
components taken into account to calculate the reduced inertia tensor, namely, the total number of particles, DM, cold baryons and hot gas. 
Rows represent different redshifts. In each plot, massive LVs ($M\geq5\times10^{12} M_\odot$ ) are shown in magenta, 
LVs with an intermediate mass ($5\times10^{11}\leq M<5\times10^{12} M_\odot$) in cyan and low-mass LVs  
($M<5\times10^{11} M_\odot$) in blue.}
\label{fig:cumhistemass}
\end{figure*}

\begin{figure*}
\includegraphics[width=13cm]{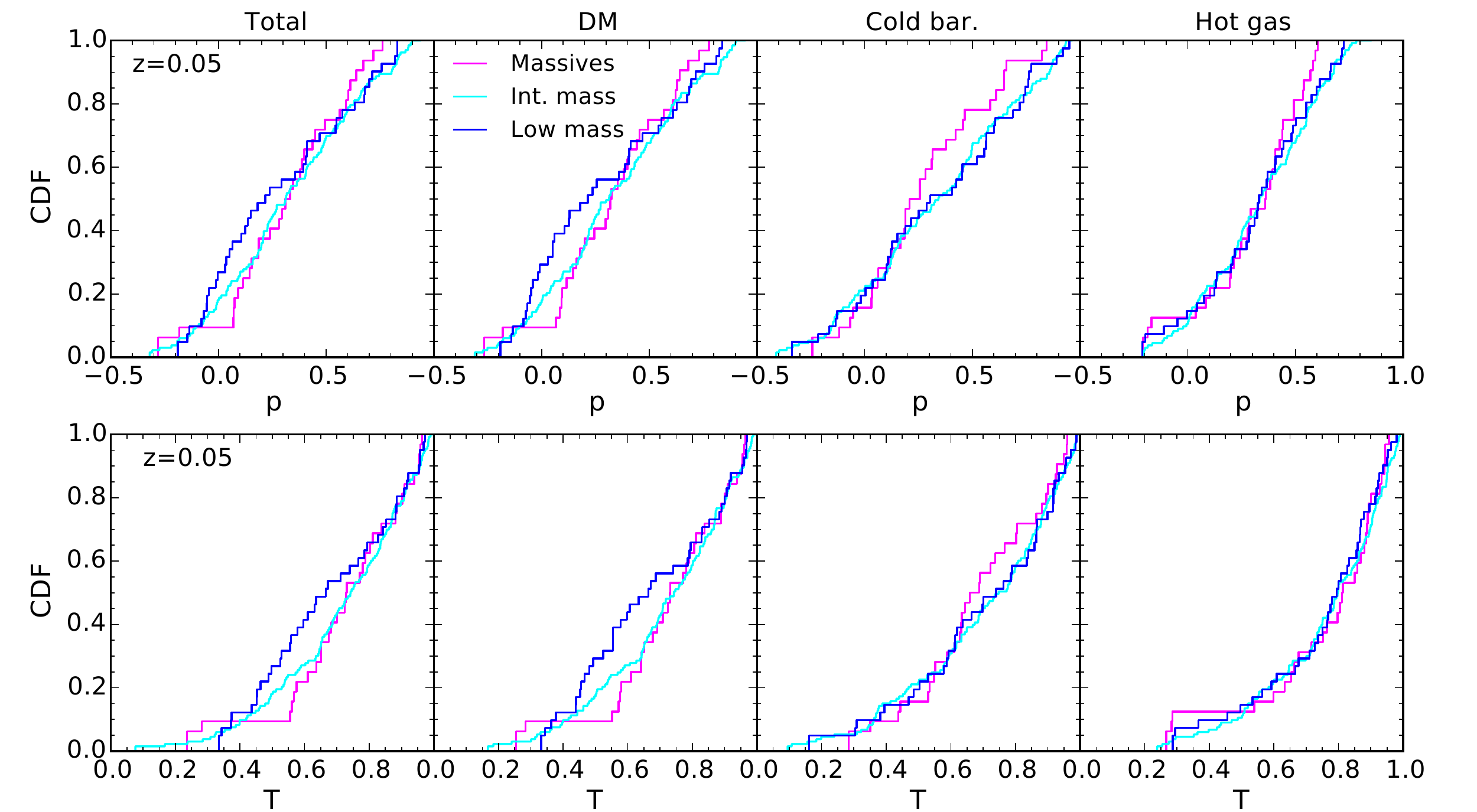}
\caption{Upper panels, CDF of the prolateness parameter, $p$. 
Lower panels, CDF of the triaxiality parameter, $T$.
From left to right the columns show the distribution binned according to the 
components taken into account to calculate the reduced inertia tensor, i.e., the total number of particles, DM, cold baryons and hot gas. 
The code colour in each plot is as in Fig.~\ref{fig:cumhistemass}}
\label{fig:cumhistpTmass}
\end{figure*}

To study the impact of the LV mass on its shape deformation, we plot in Figs~\ref{fig:cumhistemass} 
and ~\ref{fig:cumhistpTmass}, the  
CDF split by the component considered in the reduced inertia tensor calculation. From left to right we show in each column results 
obtained with all the particles, taking into account only DM particles, then cold baryons results and finally hot gas. 
Rows in Fig.~\ref{fig:cumhistemass} show cumulative probabilities at different redshifts. 
Each panel present the CDF calculated according to the binning in LV mass, massive object CDF are shown in magenta, 
intermediate mass results in cyan and low-mass CDF in blue.

In the first place, we discuss the ellipticity CDFs in Fig.~\ref{fig:cumhistemass}. As we can observe, 
the mass effects are not very relevant and moreover they almost do not evolve.
The most important mass effects appear in cold baryons at any $z$.
Indeed, the massive and low-massive LV samples at $z=3$ and $1$ 
have been determined to be drawn from different populations with the two-sample Kolmogorov--Smirnov test with $90\%$ CI; while 
the massive and intermediate mass LV samples with $95\%$ CI  at $z=3, 1$ and $0.05$.
In general, massive LVs 
tend to be more spherical across redshifts, and they have a narrower $e$ distribution than less massive ones.

In the prolateness case, the mass effects grow with time, except in the hot gaseous component. 
Hot gas independently on the mass binning is  
less spherical than the other components at $z=3$. 
At $z=1$, massive LVs are more spherical than the less massive ones for both DM and cold baryons.
The mass effect is less pronounced  in the case of hot gas. 
At $\zlow$ the tendency described above is kept (see upper panels in Fig.~\ref{fig:cumhistpTmass}).
The $p$ distribution in massive LVs is narrower than those in the other mass bins and it becomes wider faster in the low-mass bin.

Regarding triaxiality CDFs, again mass effects grow with evolution, mainly in the DM component
(see lower panels in Fig.~\ref{fig:cumhistpTmass}). 
We can also note that in both, 
the total and the DM case, there are almost no systems with $T<0.6$, specifically, there is a lack of oblate massive objects  
relative to  the other mass groups. We have tested the difference between the massive and the low-massive bins with the two-sample Kolmogorov--Smirnov  test at a $90\%$ CI. 
This mass effect is less significant in the baryon case.
Indeed, cold baryons do not present a significant mass effect, only less massive LVs tend to be more oblate than the more massive bins at $z=3$ and $1$.

Summing up, except for the hot gas  component,
 more massive LVs tend to evolve slightly more slowly from 
 their initial spherical shape than less massive ones.
This can be interpreted in terms
of the CW dynamics as follows: 
more massive objects would appear more frequently in nodes of the CW, versus less massive objects being 
present in filaments and walls. Therefore, the relative importance of anisotropic mass rearrangements
versus radial ones is lower in massive than in less massive LVs.     
Concerning the hot gas component, no relevant evolution has been detected, 
particularly after $z \sim 3$, indicating that neither the 
possible anisotropic flows towards  singularities, nor the possible pressure-induced
anisotropic outflows, have caused measurable LV mass rearrangements 
in the LV sample thereafter.

\section{Freezing-out of eigendirections and shapes }
\label{sec:Percola}

\subsection{Freezing-out times}

\begin{figure}
\includegraphics[width=8.4cm]{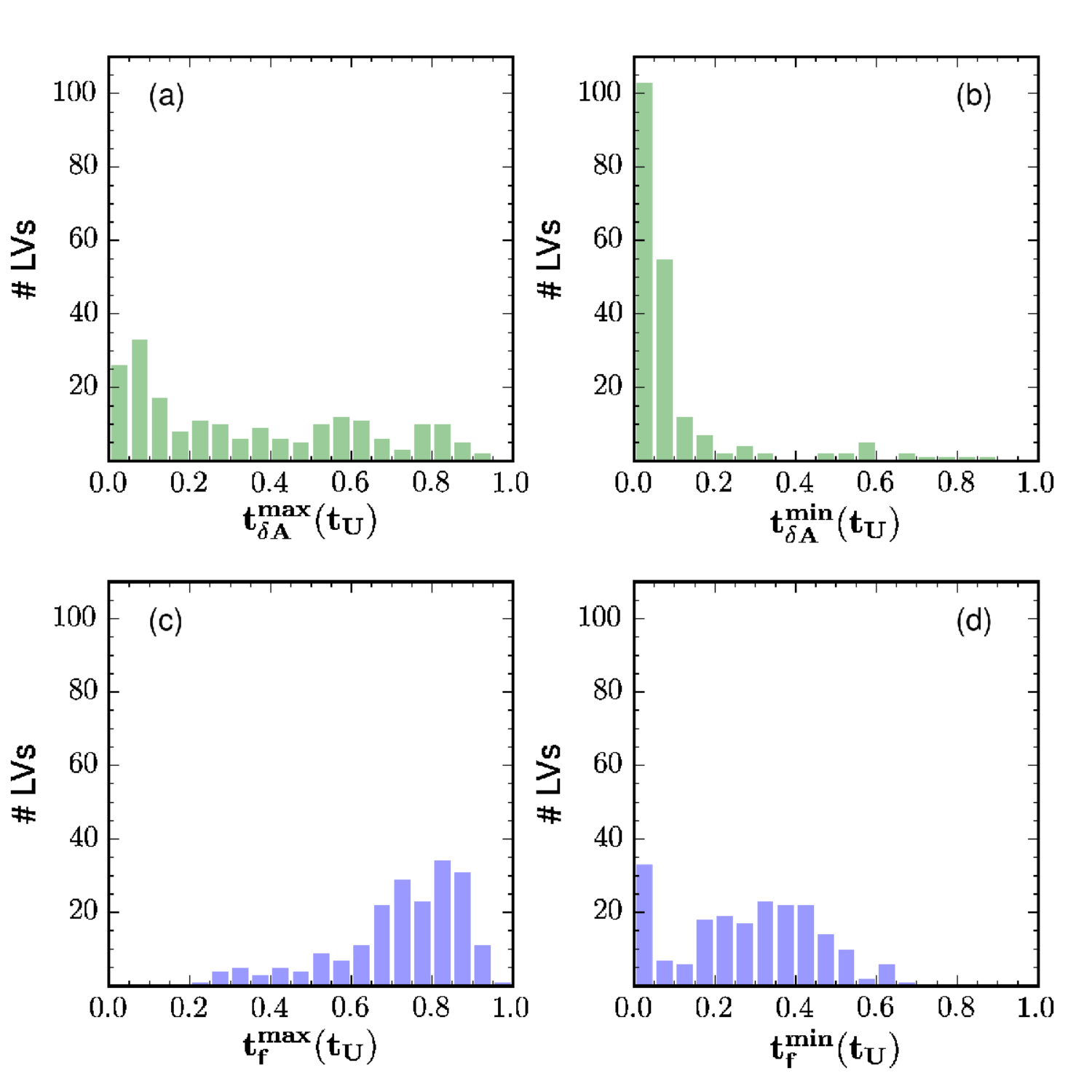}
\caption{Histograms for $\tdAmax$, $\tdAmin$, $\tfmax$ and $\tfmin$
defined with $\cos (\delta A_i) = 0.9$ and $f=0.1$.
}
\label{fig:Htmax-tmin}
\end{figure}

In the previous sections we have become aware that the $A_i(z), i=1,2$ and 3 angles evolve with time and 
$\rightarrow0$\textdegree \ before $\zlow$.
 We remind that $A_i(z)$  is the angle formed by the eigenvectors
 $ \hat{e}_i^{\rm tot}(z)$ and $\hat{e}_i^{\rm tot}(z_{\rm low})$, with $i=1,2,3$, where `$\rm tot$' stands for the eigenvectors of
the $I_{ij}^{\rm r}$ tensor corresponding to the total mass of the LV. 
Also, the evolution of the LV inertia ellipsoid declines 
in the same limit, see Figs ~\ref{fig:angAi} and ~\ref{fig:prinaxes}. 
In this section, we use the times when these eigendirections and inertia axes become frozen. 
We have calculated these freezing times to study and compare both processes and 
to look for possible mass effects.
The subject is interesting to elucidate how and when the local CW around galaxies-to-be 
becomes frozen at the scales analysed in this paper,  while it still feeds the protogalaxies at smaller scales.    

Having the $A_i(z)$ angles $\sim0$\textdegree \ during a $z$ range $z \ge \zlow$ means that the LV deformations  
become fixed in their eigendirections before $\zlow$, or, in other words,
mass rearrangements  are thereafter  organised in terms of frozen symmetry axes making
the inertia tensor diagonal, i.e.,  in terms  of a skeleton-like structure.
This motivates the search for the moment when a given LV gets its structure frozen. This is not a straightforward
issue, however, because this situation is gradually reached: all we can do is to resort to thresholds. 

In the following, we use time instead of $z$ in order to make our results clearer. 
Given a threshold angle $\delta A_i$, we define $t_{\delta A_i}$ as  the time (Universe age at the event in units of the current Universe age $t_{\rm U}$)
 when $A_i(t) \le \delta A_i$ if $t \ge t_{\delta A_i}$, (i.e., the Universe age when the $i$th eigendirection of the inertia tensor becomes fixed within
an angle $\delta A_i$). Then, we define
$\tdAmax$ and $\tdAmin$ as the maximum and minimum values of $t_{\delta A_i}, i=1,2,3$,
for each LV. That is, $\tdAmax$ for a given LV is the fractional time
 when the directions of its {\it three } eigen vectors  become frozen,
or, symbolically,   $A_i(t) \le \delta A_i$ if $t \ge \tdAmax$ 
for any direction\footnote{Note that the second and the third eigendirections become frozen at the same time.}. 
The minimum $\tdAmin$ satisfies the same condition for just one direction.
Fig.~\ref{fig:Htmax-tmin} (upper plots) 
 shows the distribution of $\tdAmax$ and $\tdAmin$ for
our sample of 206 LVs with $\delta A_i$  such that $\cos (\delta A_i) = 0.9$.

A very interesting point is to explore LV shape transformations 
relative to the freeze-out times for inertia eigendirections. 
An illustration can be found in Figs~\ref{fig:angAi} 
and ~\ref{fig:prinaxes}. Comparing both figures, we see that the principal 
axes change slightly after skeleton emergence for the particular LVs considered in this figure
by using a $10\%$ threshold (see below). 
The differences are larger for other LVs, and, indeed it is worth analysing this issue in more detail.
Therefore, to be more quantitative, we define $t_{f, a}$ as  the fractional time  when 
the inertia axis $a$ becomes frozen within a threshold
$f_a$, which is a fixed fraction of the $a(t)$ value, i.e. 
$\Delta a(t) \le  f_a $ if $t \ge t_{f, a}$,  where $\Delta a(t) \equiv \frac{\mid a(t) - a(t_{\rm low})\mid}{a(t_{\rm low})}   $.
Similarly, we define $t_{f, b}$ and  $t_{f, c}$, and then
$\tfmax$  and $\tfmin$. The former is the time when the three  inertia axes become frozen, while the latter is the time when just one axis gets 
frozen\footnote{Again, once the value of one principal axis becomes fixed, the freezing times for the other two axes are the same.}.
To have an insight of the statistical behaviour of these times, in Fig.~\ref{fig:Htmax-tmin}  (lower plots)
the histograms for $\tfmax$ and $\tfmin$ are represented for $f = 0.1$.
In this figure, right- (left-)panels correspond to the times when one (three) out of the eigenvectors or the
principal inertia axes become fixed within a $10\%$ of their final values.

\begin{figure}
\includegraphics[width=8.4cm]{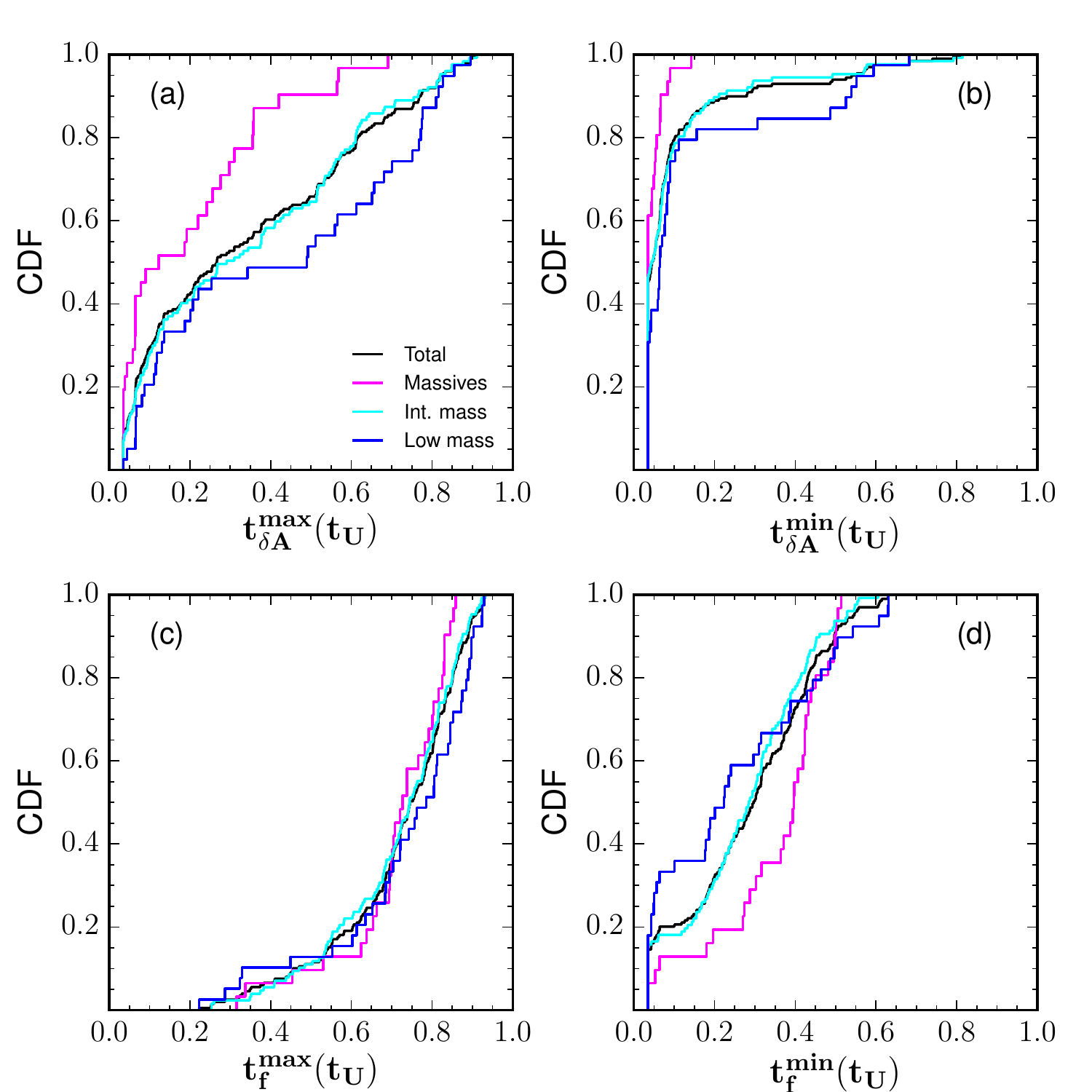}
\caption{CDFs for the same quantities in the previous figure, showing possible mass effects.}
\label{fig:CumHist_mass_effect}
\end{figure}

An interesting result is that  the time range for $\tfmax$ is narrow and late. The range of $\tdAmax$ is much
wider, which means that a high fraction of
 LVs get at high $z$ their three eigendirections fixed before the evolution of their inertia axes ends up. During  this early 
time interval, LVs change their shape with frozen symmetry axes, i.e., anisotropic matter inflows onto CW elements.
Another result is the $\tfmin$ accumulation at
the first bin of the evolution time: these are the systems having a principal axis of inertia  that keeps within 
a $10\%$ of its initial value along the evolution. They are less prolate than other
systems. 
An even higher fraction of LVs have one of their eigendirections  fixed in the first $5\%$ of the evolution
time (see Fig.~\ref{fig:Htmax-tmin}.b). 

A high fraction of  systems also got  one frozen eigendirection, while none
of their principal inertia axes is fixed yet.  However, at the end of the evolution this effect vanishes (compare Figs \ref{fig:Htmax-tmin}.b and \ref{fig:Htmax-tmin}.d). 
Finally, let us mention that LVs also spend an important  fraction of their lives with one but not three fixed eigendirections (within the thresholds used
to draw these figures, compare Figs \ref{fig:Htmax-tmin}.a and \ref{fig:Htmax-tmin}.b), 
or one but not three frozen inertia axes (compare Figs \ref{fig:Htmax-tmin}.c and \ref{fig:Htmax-tmin}.d).

\subsection{Mass effects}
\label{MEffFreez}

Next, we  look for mass effects in the distributions of $\tdAmax$ and $\tdAmin$, as well as 
in those of $\tfmax$  and $\tfmin$. This  is more clearly visualised in terms of cumulative histograms. 
In Fig.~\ref{fig:CumHist_mass_effect}, we plot the CDF for $\tdAmax$ and $\tdAmin$ 
(i.e., LV eigen  directions relative to their final values, first row) and $\tfmax$  and $\tfmin$ (principal inertia axes, second row), respectively, 
where no binning has been used. 
To analyse possible mass effects, results for the three mass groups are shown in each panel. 
The cumulative histograms in the four panels of this figure are in one-to-one correspondence with the histograms
in Fig.~\ref{fig:Htmax-tmin}.

The first outstanding result is that the time range for $\tfmax$ is roughly the same (narrow and late),
irrespectively of the mass range used (Fig~\ref{fig:CumHist_mass_effect}.c). 
This behaviour can be understood as the
consequence of $\frac{d D_{+}(t)}{d t} \rightarrow 0$ at late times, a global effect
causing anisotropic flows to vanish, 
see $\S$\ref{ZAImpli} for more details. 
Nevertheless, there exists a mass effect in $\tdAmax$  (Fig.~\ref{fig:CumHist_mass_effect}.a), 
with the least-massive LVs showing a delay in the spine emergence or in getting their three eigendirections frozen 
with respect to more massive ones, the differences being
more marked at early times. This is somewhat expected from the previous 
discussion on the effects of the eigenvalue landscape 
heights on the timing of spine emergence, in $\S$\ref{ZAImpli}.

Fig.~\ref{fig:CumHist_mass_effect}.b exhibits strong mass effects too. Indeed, at early times the most massive systems  
get one out of their three eigendirections frozen sooner than less massive ones. 
In fact, $\sim$ $95\%$ of the massive LV subsample has one of their eigendirections
fixed at $t/t_{\rm U} \simeq 0.1$. 
This mass segregation can be understood in the light of the considerations made in  $\S$\ref{ZAImpli}, where
we concluded that the first CW elements tend to appear and percolate earlier on  within massive LVs than within less massive ones.

On the other hand, the freezing-out times for the principal axis of inertia (panel \ref{fig:CumHist_mass_effect}.d) display a 
remarkable mass effect, although just at early times. 
Later on, irrespective of their mass, 
no LV gets its first 
principal axis of inertia
  fixed later than $t/t_{\rm U} \simeq 0.55$.   This upper bound on $\tfmin$ might be a consequence of both, 
the $\frac{d D_{+}(t)}{dt} \rightarrow 0$ after the $\Lambda$ term dominates the 
Universe expansion, and the fact that flows towards
walls are the first to vanish at a local level. 
The mass effect lies in massive systems having 
 their $\tfmin$ delayed at early times in relation to less massive ones
(consistently with what was found in $\S$ \ref{Shape-Evol}), the difference vanishing 
at $z \sim 1$.

\begin{figure}
\includegraphics[width=8.4cm]{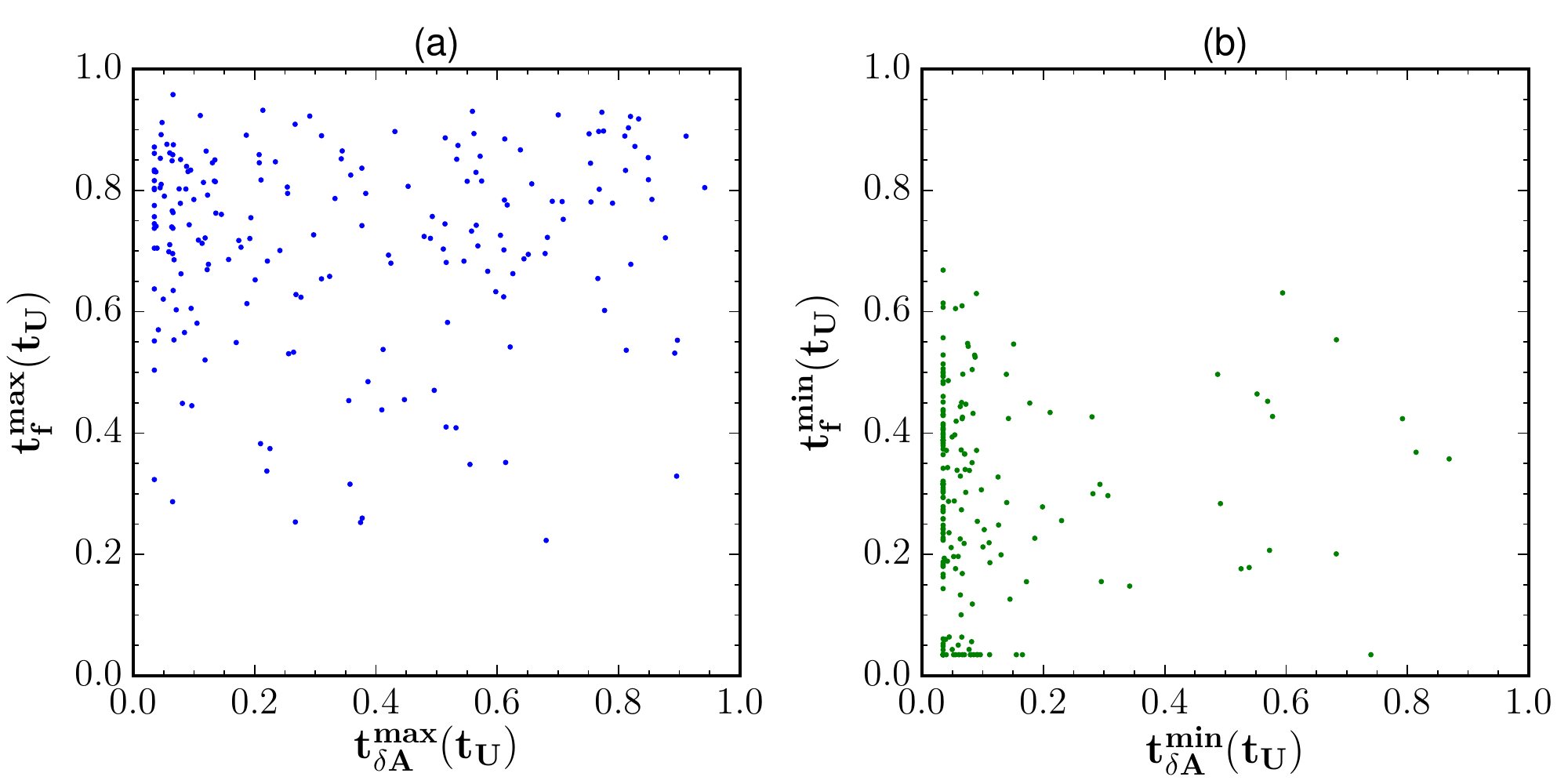}
\caption{Scatter plots of $\tfmax$ versus $\tdAmax$ (left) and $\tfmin$ versus $\tdAmin$ (right).
}
\label{fig:tmax-tminPlot}
\end{figure}

Finally, to look for correlations, the $\tfmax$  and $\tfmin$ for our
sample of LVs are plotted versus their respective  $\tdAmax$ and $\tdAmin$,
in Fig.~\ref{fig:tmax-tminPlot} for $f=0.1$ and $\cos(\delta A_i) = 0.9$.
No outstanding correlation exists in any case, but we see that indeed, most systems have their eigendirections
fixed before their principal axes got frozen.

Summing up, we observe that on average  eigendirections (either one or the three) for massive LVs  become 
fixed at earlier stages than that 
of less massive LVs. Nevertheless, no relevant mass effects are found for principal inertia axis freezing times.
In addition, eigendirections become in general fixed before mass flows onto the corresponding CW elements stops,
the time delay being particularly long for the first eigendirection relative to the first principal axis
in massive systems.
Thus,  the first eigendirection in massive systems gets fixed quite a while before the accretion onto it  stops.

\section{Discussion: Possible Scale Effects}
\label{subsec:scaleeffects}

In  Section \ref{subsec:methods}, when describing how to build up the LV sample, 
a value of $R_{\rm high} = K\times r_{\rm vir, low}$ with $K = 10$ 
has been chosen to define the LV at $\zhigh$. 
As explained there, this choice was motivated as a compromise between low $K$ values, ensuring a higher number of LVs in the sample, 
and a high $K$, ensuring LVs with  high enough number of particles so that we obtain meaningful LVs. 
However, $K = 10$ is by no means the unique value that satisfies these constraints. 
Therefore, it is important to test out the possible effects of changing this value under the same constraints.

To this aim, we have repeated all the calculation using $K = 7.5$ and $15$. 
The LV building up (see section 2.2) has been repeated with the same SKID identified haloes at $\zlow$ as first step. 
Nonetheless, when $K = 15$ is used, some of the LVs do not satisfy anymore the
condition of having all their particles inside the hydrodynamic zoomed volume.
These particular LVs have been removed from the initial sample of 206 LVs, in such a way 
that we are finally left with 159 LVs for $K = 15$. 
This problem does not exist when using $K = 7.5$; however, 
to probe the scale effects, we need samples that contain the same $\zlow$ 
SKID-identified haloes as starting point in the three scales. Therefore, only these 159  well-behaved 
LVs (a subset of the initial $K = 10$ sample) have been used to analyse the scale effects. 

The first relevant outcome is that there is no substantial difference when 
results obtained with the subsample of 159 LVs and with the sample used along this paper (206 LVs) for $K = 10$
are compared. 

In the following subsections, we will compare the results obtained with each of the three samples of 159 LVs, dubbed 
according to its $K$ value, $\kshort$, $\kinterm$ and $\klarge$.

\subsection{Effects on eigenvector orientation evolution}

\begin{figure}
\centering
\includegraphics[width=7.5cm]{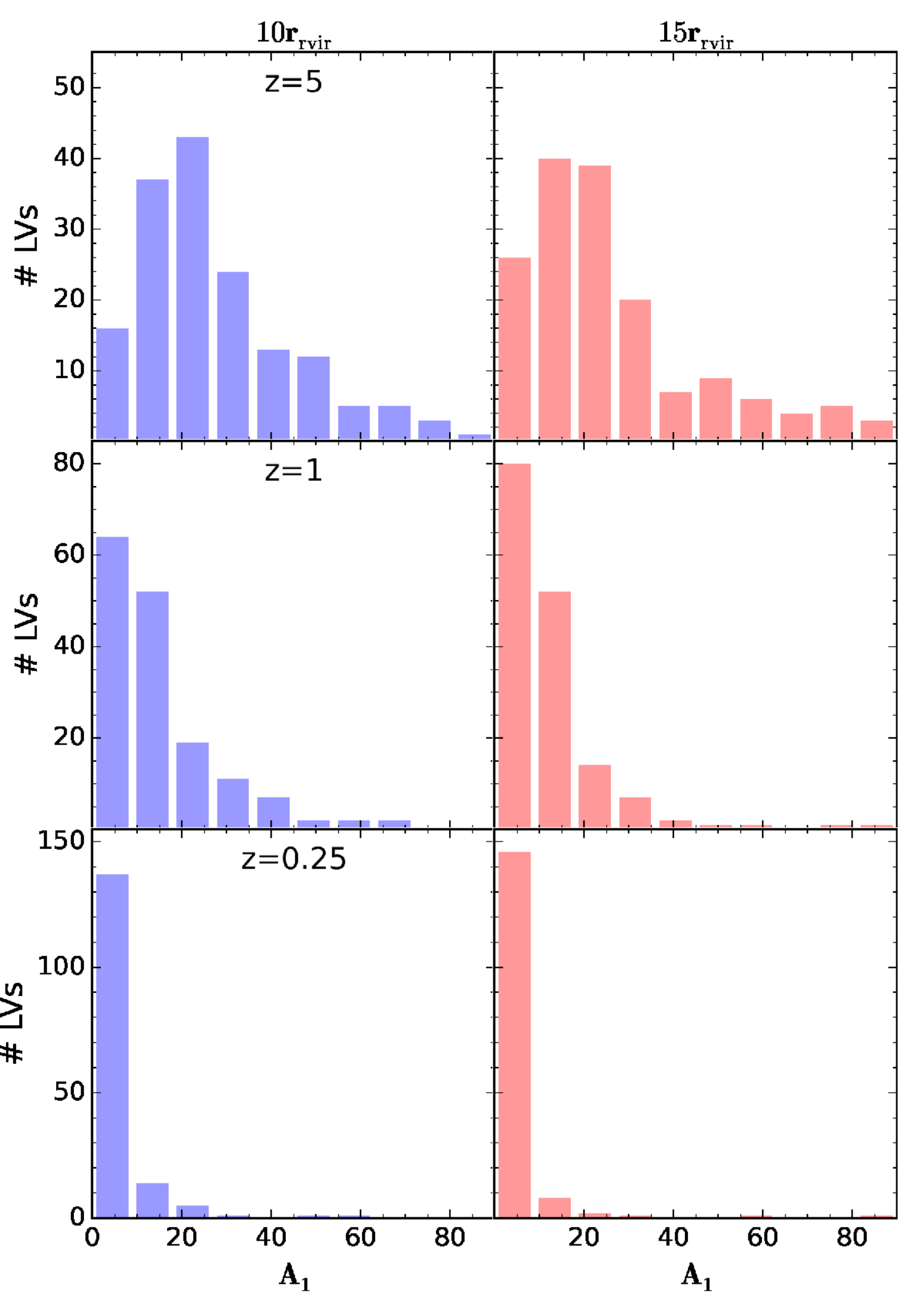}
\caption{Histograms of the $A_1$ distribution at different redshifts for the $\kinterm$ and $\klarge$ samples (left- and 
right-hand columns, respectively).
}
\label{fig:histangAi_scale}
\end{figure}

Concerning the evolution across redshifts of the $I_{ij}^{\rm r}$ eigendirections relative to
their final values at $z_{\rm low}$ (Fig.~\ref{fig:direc-evol}), no relevant differences
have been found between the histograms obtained with the $\klarge$ and $\kinterm$
samples at the same redshifts. Fig.~\ref{fig:histangAi_scale} illustrates this behaviour,
showing that the $A_1$ angle distributions for $\klarge$ are similar to those found
with $\kinterm$ at different $z$ pairs, see $\S$\ref{sec:EscFree} for more details.

In addition, no scale effects appear in the angles formed by  the 
 eigenvectors, $\hat{e}_i^{\rm tot}(z)$, $i=1,2,3$,  arising  from the overall
 matter distribution with the same eigenvectors calculated with the different components
(i.e., those angles whose distribution for the sample of 206 objects is given in Fig.~\ref{fig:histangei}.)

\subsection{Effects on shape evolution}

To gain further insight, the 159 LV subsample  has been split according to the LV masses. 
In order to assure that we are comparing the same mass bins for the three scales, 
we have mapped the LVs belonging to the three mass ranges defined for the $\kinterm$ sample to 
the LVs of the  $\klarge$ and  $\kshort$ scales.

\begin{figure}
\includegraphics[width=8.4cm]{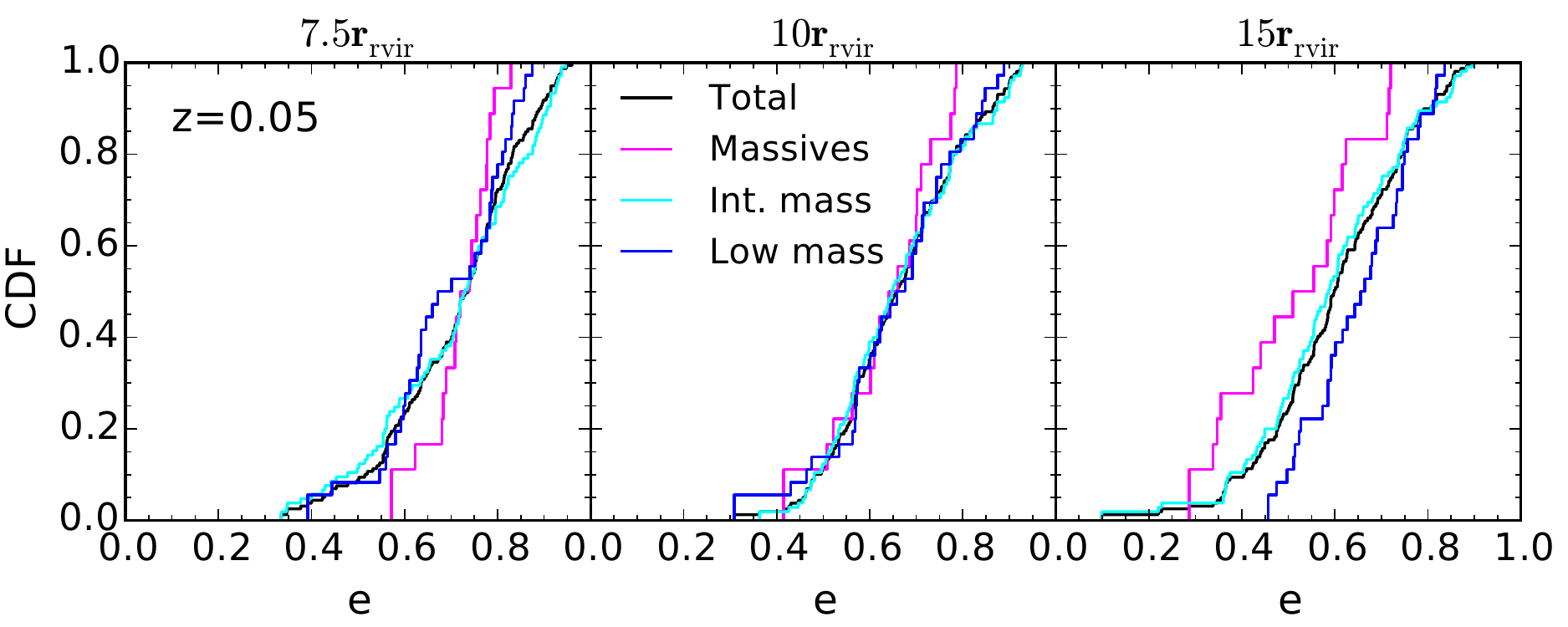}
\caption{CDFs of the ellipticity  at $\zlow$ portraying mass effects obtained with  the three different scales, $\kshort$
$\kinterm$ and $\klarge$. 
}
\label{fig:shape_scale}
\end{figure}

Important results concerning shape evolution are as follows.
\begin{enumerate}
\item 
No relevant differences in the evolution patterns have been found 
between the least massive LV group ($M<5\times10^{11} M_\odot$ in the $\kinterm$ sample) 
when followed in the $\klarge$, $\kinterm$ and $\kshort$ samples (see Fig.~\ref{fig:shape_scale}, blue lines).
That is, these LVs are hardly sensitive to the $K$ scale in their evolution.
The scale effects are only slight  between the $\klarge$ and $\kinterm$ samples 
when no mass splitting in the LV sample is  performed (see Fig.~\ref{fig:shape_scale}, black lines).

\item 
LVs in the massive group are sensitive to the $K$ scale, with the $\kshort$ samples showing particular
differences.
 Fig.~\ref{fig:shape_scale} is an example of such a behaviour, likely due
to the wall effect, whose formation is better sampled with $\klarge$.
Also, walls are more frequent in
 massive LVs. See $\S$\ref{sec:EscFree} for more details.

\item In any case, the qualitative results reached in $\S$  \ref{sec:results}  
about component effects in shape deformations are stable when comparing $\klarge$ and $\kinterm$ samples. 

\end{enumerate}

\begin{figure}
\includegraphics[width=8.4cm]{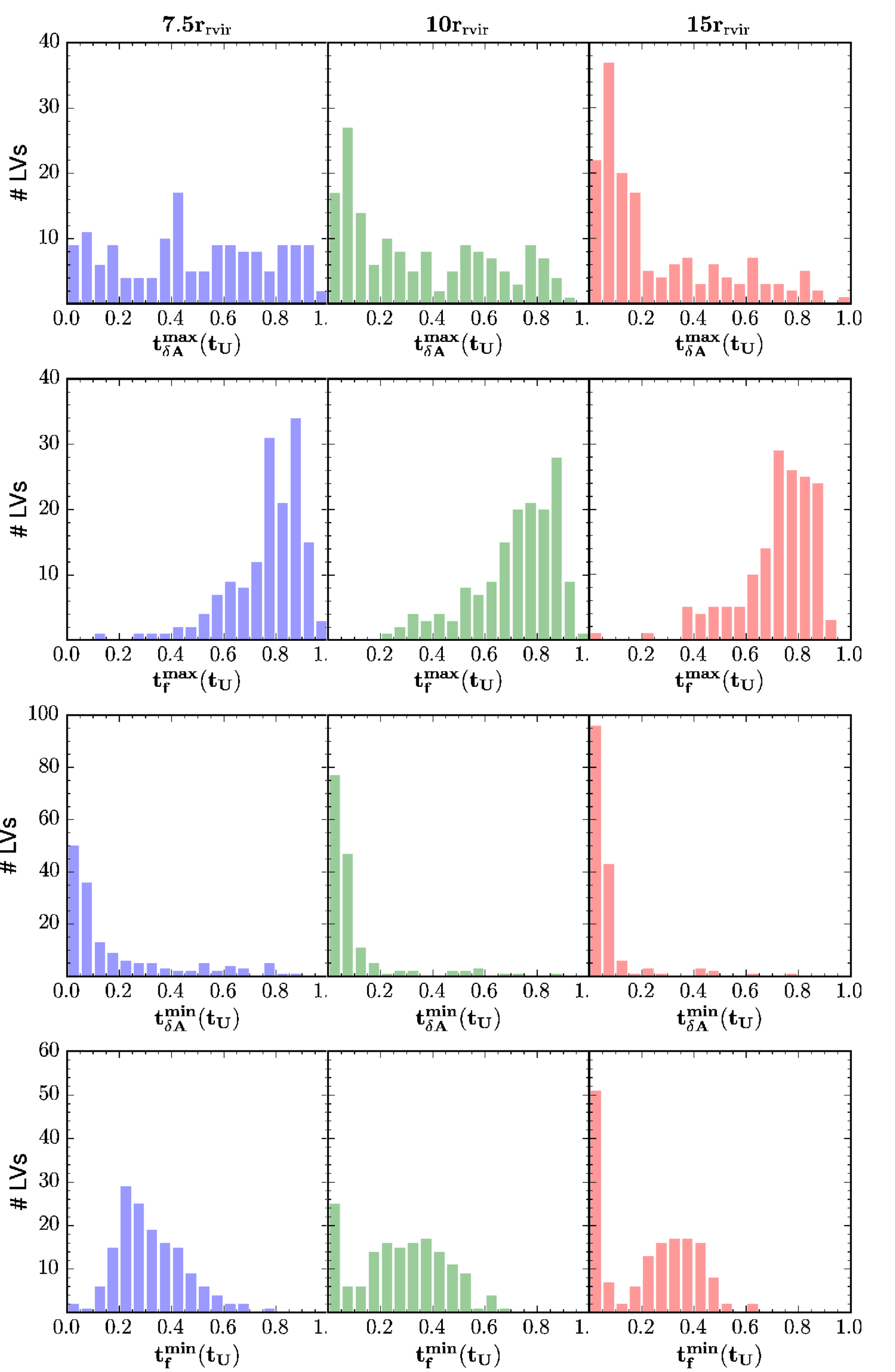}
\caption{Histograms for $\tdAmax$, $\tdAmin$, $\tfmax$ and $\tfmin$
defined with $\cos (\delta A_i) = 0.9$ and $f=0.1$. Columns show the results obtained for the 
three samples, $\kshort$ $\kinterm$ and $\klarge$.
}
\label{fig:freezingout_scale}
\end{figure}

\begin{figure}
\includegraphics[width=8.4cm]{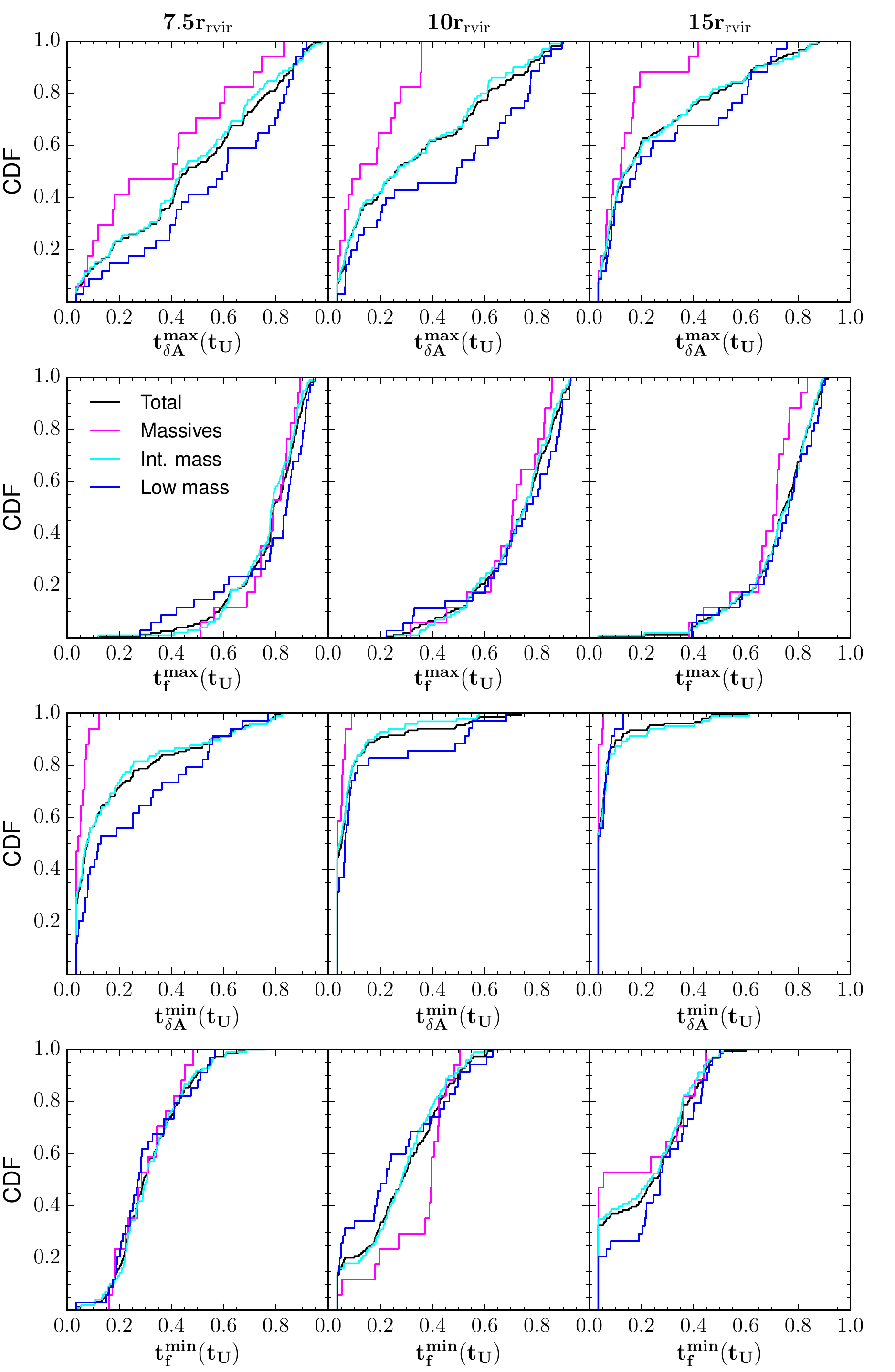}
\caption{CDFs of  $\tdAmax$, $\tdAmin$, $\tfmax$ and $\tfmin$ at different $K$ scales.
}
\label{fig:freezingout_masseff_scale}
\end{figure}

\subsection{Effects on freezing-out times}
\label{sec:EscFree}

Fig.~\ref{fig:freezingout_scale} shows the histograms for the $\tdAmax$,  $\tfmax$, $\tdAmin$, and $\tfmin$ 
times for samples using different $K$ scales.
It is clear from this figure that 
while the results for the $\klarge$ and $\kinterm$ samples are 
roughly consistent with each other, those for the $\kshort$ sample differ.
The only exception is the $t_{f}^{\rm max}$ time distribution (second row), whose pattern is the same at any scale,
 namely rather late and peaked. 
Recall that  $t_{f}^{\rm max}$ is the time when the three inertia axes are fixed to within $10\%$ of their final values, i.e., the time when all anisotropic fluxes stop. This behaviour can be understood as the
consequence of $\frac{d D_{+}(t)}{d t} \rightarrow 0$ at late times, that is a global effect.

A key point to understand some aspects of Fig.~\ref{fig:freezingout_scale} behaviour, 
is the fact that the $\kshort$ scale is too short to suitably sample the whole process of wall formation within some  LVs.
As a consequence, since the first flows to vanish are those towards walls (see $\S$ ~\ref{ZAImpli}),
the $t_{f}^{\rm min}$ time (when the first inertia axis is fixed to within $10\%$ of its final value)
will be delayed at high $z$ in the $\kshort$ sample, as observed in Fig.~\ref{fig:freezingout_scale}, fourth row.
A remarkable results is that, 
irrespective of the $K$ scale,  
no LV has its first inertia axis frozen later than $t/t_{\rm U} \simeq 0.55$.
This result reinforces our interpretation given  in $\S$ \ref{MEffFreez} that this effect  is,
at least partially, a consequence of the $\frac{d D_{+}(t)}{dt} \rightarrow 0$ tendency at latter times.

The process of wall formation could be also  the reason of the similarities and differences 
 found in  the distributions of the $t_{\delta A}^{\rm min}$ times (when the first eigenvector
direction is fixed to within a $10\%$).
The panels of the third row of Fig.~\ref{fig:freezingout_scale} 
show that their distributions are always peaked
towards very early times, meaning that the $\hat{e}_3$ eigenvector of the $I^{r}_{ij}$ for some LVs
freezes its direction very early,  following wall formation. 
In addition, we see that as we move from $\klarge$ to $\kinterm$  to $\kshort$,
a delay appears, not so relevant between the $\klarge$ and $\kinterm$ samples.
Again, this can be interpreted in terms of the inadequacy of the shorter scale to
properly catch the characteristics of wall formation in some LVs.

Finally, we address the scale effects on  $t_{\delta A}^{\rm max}$
(first row of Fig.~\ref{fig:freezingout_scale}). These are the times when the LV
 orientations become frozen to within $10\%$ of their final values, i.e., the times
marking the skeleton emergence locally within each LV.
While its distribution is rather peaked at early times for both, the $\klarge$ and the $\kinterm$ samples, 
it flattens as we go to $\kshort$. 
Once again, the poor wall formation sampling in most $\kshort$ LVs is likely to be 
the cause of this difference.

It is worth noting that the qualitative features found in $\S$ \ref{sec:Percola} are stable 
under the change in $K$. 
For instance, mass effects can be analysed from Fig.~\ref{fig:freezingout_masseff_scale},
 where we show 
the $\tdAmax$,  $\tfmax$, $\tdAmin$ and $\tfmin$ mass-binned CDFs  
(first, second, third and fourth rows, respectively), at different scales (columns).
Then, we can note that, regardless of the $K$ value,  
the $\tdAmax$ and $\tdAmin$ distributions show qualitatively similar mass effects,  with 
the most massive LV group fixing either one or their three eigenvalues earlier on
than LVs in the intermediate or less massive group (as expected).
Moreover, the $\tfmax$ distribution does not show relevant mass effects whatever the considered scale.
Finally, irrespective of the scale, the $\tfmin$ distributions do not show relevant mass effects after $t/t_{\rm U} \simeq 0.4$,
as expected from the previous analyses. 
At low $z$, some mass segregation is found, and
furthermore, it qualitatively depends on the scale. 
This is the only one exception to the stability under the change in $K$.
These results could  reflect the difficulty of catching the end of the mass flows in only one direction
when the contribution of wall formation is combined with mass effects.

Summing up, the differences in the freezing-out times are not very relevant 
when using the $\klarge$ or the $\kinterm$ samples.
Their distributions show similar patterns, in particular when mass effects are considered.

\section{Summary and conclusions}
\label{sec:conclusions}

In this paper, we present a detailed analysis of the local evolution of 206 Lagrangian Volumes (LVs) selected at high redshift around proto-galaxies. 
These galaxies have been identified at $\zlow =0.05$ in a large-volume hydrodynamical simulation
run in a $\Lambda$CDM cosmological context and they have a mass range $1 - 1500 \times 10^{10} M_\odot$.
We follow the dynamical evolution of the density field inside 
these initially spherical LVs from $\zhigh=10$ up to $\zlow=0.05$,
witnessing mass rearrangements within them, leading to the emergence of a highly anisotropic,
complex, hierarchical organisation, i.e., the {\it local} cosmic web (CW).
Indeed, at $\zlow$ LVs acquire overall anisotropic  shapes as a consequence of mass inflows
onto singularities along cosmic evolution, in  such a way that 
some relevant aspects of these mass arrangements can be described 
in terms of the reduced inertia tensor $I_{ij}^r$ evolution, as given by its principal directions  and inertia axes, $ a \ge b \ge c$.

Our analysis focuses on the evolution of the principal axes of inertia and their corresponding eigendirections,
paying particular attention to the times when the evolution of these two structural elements declines.
In addition, mass and component effects (either DM, cold or hot baryons) along this process have also been investigated. 

In broad terms, we have found that local LV evolution follows the predictions of the Zeldovich Approximation \citep[ZA, ][]{Zeldovich:1970} and 
the Adhesion Model \citep[AM, ][]{Gurbatov:1984,Gurbatov:1989,Shandarin:1989,Gurbatov:1991,Vergassola:1994} 
when both caustic dressing \citep{Dominguez:2000} and mutual gas versus CW effects \citep[see Section 3 and][]{DT:2011,Metuki:2014} are taken into account. 
Evolution also entails baryon transformation into stars inside the densest regions of the web and 
gravitational gas heating following the collapse.
More specifically, these are our main results.

Dark matter dominates dynamically the LV shape deformations over the baryonic component, as expected from 
hierarchical structure formation.   
Deformations transform most of the initially spherical LVs into prolate shapes, i.e. filamentary structures, 
in good agreement with previous findings \citep{AragonCalvoa:2010,Cautun:2014}. Cold baryons follow DM behaviour in general,
 but with  some departures from it,  departures  that  rise  as evolution proceeds. Accordingly, the
number of LVs having their cold baryonic principal axes in directions that differ  
from  the ones calculated with their DM content is negligible at
$\zhigh$, and it keeps low along the evolution, but increases with time ($\sim 25\%$ at $\zlow$). 
On the contrary, the hot gas eigendirections have a flatter distribution  at $\zhigh$ and then they tend to converge to those calculated with DM. However, 
only $\sim$ half of them reach such convergence at $\zlow$. 
This tendency towards convergence is due to the fact that the hot gaseous component traces the locations where 
sticking events, in particular filament and node formation, have taken place. 
The mass fraction involved in these processes increases with evolution, and consequently we expect a tendency of the hot gas to be aligned with the total eigendirections.

In terms of shape evolution, a clear component effect has  been found regarding  the way how the evolution occurs. 
In fact, hot gas shapes 
do not exhibit important evolution because, as said above,
gravitationally heated gas marks out  the places where sticking events have taken place, and because, in addition, 
no evidence for important anisotropic mass rearrangements in this component have been found in this paper.
 The only remarkable effect is that the needle-like or flat shapes
shown by hot gas in some LVs around $z=5$, are transformed  at lower $z$s. 
As mentioned before, DM and cold baryons shapes do evolve, with cold baryons achieving an even more
pronounced filamentary structure than DM ones as a consequence of dissipation.   
Additionally, some mass effects have also been found in the generic evolution of shapes,   
with lower mass  LVs evolving  towards more pronounced filamentary structures on average 
and earlier on than the more massive ones.

A remarkable result of our analyses is that the evolution of LV deformations declines.  This means that both  
the LV eigendirections, as well as their principal axes of inertia ($a,b$ and $c$) values become roughly
constant before $\zlow$. 
This is a smooth effect that can be only defined in terms of thresholds.
Taking a $10\%$ of the final values, shape (i.e., $a,b$ and $c$ values) freezing-out time distribution 
has a narrow peak ($\sim 0.2$ at each side) around $t/t_{\rm U}=0.8$. This happens later than the freezing-out times for the three 
LV eigendirections, whose distribution  peaks around $t/t_{\rm U}=0.1$ and 
 then it is flat until $t/t_{\rm U} \sim 0.8$ when it decays.

By plotting individual freezing times for shapes and eigendirections, respectively (see Fig.~\ref{fig:tmax-tminPlot}.a), we note that first, most of  the LVs
fix their three axes of symmetry (like a skeleton), 
and later on their shapes are fixed. This result is in good agreement with 
 \citet{vanHaarlem:1993,vandeWeygaert:2008,Cautun:2014} and \citet{Hidding:2014} findings. Moreover, the ZA and the AM predict 
that walls, filaments and nodes undergo mass flows from underdense regions 
 to denser environments, that continue after skeleton emergence.

As a general consideration, it has been found that mass rearrangements at the scales taken into account have always been highly anisotropic. 
Therefore, the mass streaming towards walls and filaments 
has  been extremely anisotropic, and, to a lesser extent, towards nodes as well. 
 In particular, galaxy systems form in environments that  have a rigid spine at scales 
of a few Mpc, from whose skeleton a high fraction of mass elements that feed protogalaxies are collected. 

Due to anisotropic mass accretion, it turns out that in general the direction of just {\it one} of the LV eigen vectors or
the value of {\it one} of their axes get frozen while the other two still continue changing. Again, for each LV there is a time
delay between the moment when the first of its eigendirections get fixed (happening within the first $20\%$ of the Universe age) 
and the moment when the value of one of its principal axes becomes constant (peaking around $t/t_{\rm U}=0.35$).
Therefore, we again find a situation where first the flow direction is fixed (as a first piece in the skeleton
emergence)  while the mass flows persist.

Even more interesting because of its possible astrophysical implications (see discussion below) is our finding that more massive 
LVs fix their skeleton earlier on than less massive ones, either considering just one or the three
eigendirections. These results are not surprising since the dynamical processes involved in the spine emergence are 
faster around massive potential wells.

Concerning shape transformation decline, there are no relevant mass effects as far as
the complete shape freezing-out is considered. When just one axis value is taken into account, however, an early delay of more 
massive LVs compared to less massive ones clearly stands out, delay that vanishes at half of 
the Universe age.  

When building up the LV sample at $z_{\rm high}$ a value of 
$R_{\rm high} = K\times r_{\rm vir, low}$ with $K  = 10$ 
has been used to define the LV at this redshift.
This choice was motivated as a compromise between low $K$ values, ensuring a higher number of LVs in the sample, 
and a high $K$, ensuring that LVs are  large enough to meaningfully  sample the CW emergence around forming galaxies. 
As this $K = 10$ value is not the unique value satisfying these constraints,
the complete analysis has been repeated using $K  = 7.5$ and $15$ instead.
We have found that 
when using the $K = 15$ or the $K = 10$ samples, no relevant 
differences  in the LV eigenvector orientations,  
shape deformations and freezing-out times appear.
Therefore, using $K  = 10$ is in a sense the best choice.

It is important to remark that no explicit feedback has been implemented in the simulations
analysed here, but SF regulation through the values of
the SFR parameters. We remark that  the issues discussed
in this paper entail considerably larger characteristic scales than the ones related to stellar feedback. Hence,   
it is unlike that the details of the star formation rate, and those of stellar feedback in particular,
could substantially alter the conclusions of this paper, at least at a qualitative level.
Concerning the inner halo scale, we recall that to properly explore the impact of SNe feedback into filamentary patterns,
high enough resolution in order to resolve SNe remnants into the Taylor--Sedov phase are needed.
Such simulations are available (the NUT simulations, at sub-parsec scale),
but only up to $z=9$ \citep{Powell:2011}. Therefore, we still  have to wait to properly understand  how SNE feedback can possibly affect 
the CW emergence and dynamics.  However, the findings so far, at high $z$, suggest that the filamentary patterns
are  essentially untouched by SNe feedback \citep{Powell:2013}.

\subsection{Astrophysical Implications}

The results summarised so far could have important implications
in our understanding of galaxy mass assembly, raising different interesting issues.

According to our results, it takes longer for less massive systems to fix their spine, possibly
making it easier for these systems  to acquire  angular momentum  through filament transverse motions relative  to the galaxy haloes. 
In fact, recent studies on galaxy formation  \citep{Kimm:2011,Pichon:2011,Tillson:2012,Dubois:2014} in the CW context,
underline the role that filament motions  in the protogalaxy environment could have had in endowing filaments, and eventually
the adult galaxy, with angular momentum.
If real, this effect could contribute to the mass-morphology correlation
\citep[see for instance][]{Kauffmann:2003}.

Our results also point towards (major) mergers events having a high probability to occur within filaments.
This is an important issue, though beyond the scope of this paper. In fact, if confirmed, this could decrease 
 the allowed merger orbital parameter values \citep[see for example,][]{Lotz:2010,Barnes:2011}, as most mergers would have
these parameters constrained within the filament.

Another issue concerns the use of close pairs in merger rate calculations from
observational data, under the hypothesis that these systems are bound and about to merge
\citep[see, for instance][]{Patton:2000,Bell:2006,Kartaltepe:2007,Patton:2008,Robaina:2010,Tasca:2014,LopezSanjuan:2014}. In this respect, some interesting efforts have been made to correct the statistics of
 pairs that are close in angular distance
from chance superposition effects on the line of sight, \citep[see e.g.,][]{Kitzbichler:2008,Patton:2008},
whose results are used by other authors in this field.
Our results reinforce the need for these analyses,
in the sense that a detailed determination of these 
corrections, including their dependence on
the galaxy properties, merger parameters and environment,  
could be  crucial for a more elaborated understanding of the
relationship among close pair statistics and merger rates.

Finally we very briefly address the question of the warm-hot  gas distribution at intermediate
scales. Our results point to the web structure being marked out by hot gas from high redshifts. 
Indeed, at scales of $4-8$ Mpc and at $\zlow$, hot gas traces the CW elements.
Note that there is observational evidence of warm-hot gas at large scales in a filament joining
Abell clusters A222 and A223 \citep{werner:2008}, where the DM component has also been 
detected \citep{dietrich:2012}, and more recently 
preliminary evidence of hot gas in cluster pairs has been found  
 from the redMaPPer catalogue 
\citep{Rykoff:2014} along  the sightline of a QSO by \citet{Tejos:2014b}, 
(see also his presentation in The Zeldovich Universe, Genesis and Growth of the Cosmic Web, 2014, IAU Symposium). 
Our results concern smaller scale structures, and they indicate that hot gas traces the
CW since the moment when gas is heated at high redshift. Indeed, 
hot gas maps out the sites where the most violent dynamical events 
have occurred, such as filament, and, more particularly, node formation. 
Confirming warm-hot gas in filaments at different scales is a major challenge for the advance 
of our understanding of
galaxy formation \citep[see for example][for details]{Kaastra:2013}.

\section*{Acknowledgements}
We thank  Arturo Serna
for allowing us to use results of simulations.
We thankfully acknowledge to D. Vicente and J. Naranjo for the assistance and technical expertise
provided at
the Barcelona Supercomputing Centre, as well as the computer resources provided by BSC/RES (Spain).
We thank DEISA Extreme Computing Initiative (DECI) for the CPU time allowed to GALFOBS project.
The Centro de Computaci\'on Cientif\'ica (UAM, Spain) has also provided computing facilities.
This investigation was partially supported by the MICINN and MINECO (Spain) through the grants
AYA2009-12792-C03-02 and AYA2012-31101 from the PNAyA, as well as by
the regional Madrid V PRICIT programme through the ASTROMADRID network
(CAM S2009/ESP-1496) and the `Supercomputaci\'on y e-Ciencia'
Consolider-Ingenio CSD2007-0050 project.
 SR thanks the MICINN and MINECO (Spain) for financial
support through an FPU fellowship.

\bibliographystyle{mn2e}

\bibliography{sandrarobles}

\label{lastpage}

\end{document}